\documentclass[sigconf,authorversion]{acmart}

\usepackage{multirow}
\usepackage{graphicx}
\usepackage{subfig}
\usepackage{enumitem}
\AtBeginDocument{%
  }

\copyrightyear{2025}
\acmYear{2025}
\setcopyright{acmlicensed}

\acmConference[CIKM '25] {Proceedings of the 34th ACM International Conference on Information and Knowledge Management}{ November 10--14, 2025}{Seoul, Republic of Korea.}
\acmBooktitle{Proceedings of the 34th ACM International Conference on Information and Knowledge Management (CIKM '25), November 10--14, 2025, Seoul, Republic of Korea}
\acmISBN{979-8-4007-2040-6/2025/11}
\acmDOI{10.1145/3746252.3761004}

\graphicspath{{./figs/}} 

\begin{document}

\title[Improving Recommendation Fairness via Graph Structure and Representation Augmentation]{Improving Recommendation Fairness via Graph Structure\\ and Representation Augmentation}


\author{Tongxin Xu}
\authornote{Both authors contributed equally to this research.}
\affiliation{%
  \institution{Guilin University of Electronic Technology}
  \city{Guilin}
  \state{}
  \country{China}}
\email{xutoncy@gmail.com}

\author{Wenqiang Liu}
\authornotemark[1]
\affiliation{%
  \institution{Guilin University of Electronic Technology}
  \city{Guilin}
  \state{}
  \country{China}}
\email{liuwenqiang0927@gmail.com}

\author{Chenzhong Bin}
\authornote{Corresponding author}
\affiliation{%
  \institution{Guilin University of Electronic Technology}
  \city{Guilin}
  \state{}
  \country{China}}
\email{binchenzhong@guet.edu.cn}

\author{Cihan Xiao}
\affiliation{%
  \institution{Johns Hopkins University}
  \city{Baltimore}
  \state{}
  \country{United States}}
\email{cxiao7@jhu.edu}

\author{Zhixin Zeng}
\affiliation{%
  \institution{Guilin University of Electronic Technology}
  \city{Guilin}
  \state{}
  \country{China}}
\email{zxzeng@guet.edu.cn}

\author{Tianlong Gu}
\affiliation{%
  \institution{Jinan University}
  \city{Guangzhou}
  \state{}
  \country{China}}
\email{gutianlong@jnu.edu.cn}

\renewcommand{\shortauthors}{Tongxin Xu et al.}
\settopmatter{printacmref=true}

\begin{abstract}
Graph Convolutional Networks (GCNs) have become increasingly popular in recommendation systems. However, recent studies have shown that GCN-based models will cause sensitive information to disseminate widely in the graph structure, amplifying data bias and raising fairness concerns. While various fairness methods have been proposed, most of them neglect the impact of biased data on representation learning, which results in limited fairness improvement. Moreover, some studies have focused on constructing fair and balanced data distributions through data augmentation, but these methods significantly reduce utility due to disruption of user preferences. In this paper, we aim to design a fair recommendation method from the perspective of data augmentation to improve fairness while preserving recommendation utility. To achieve fairness-aware data augmentation with minimal disruption to user preferences, we propose two prior hypotheses. The first hypothesis identifies sensitive interactions by comparing outcomes of performance-oriented and fairness-aware recommendations, while the second one focuses on detecting sensitive features by analyzing feature similarities between biased and debiased representations. Then, we propose a dual data augmentation framework for fair recommendation, which includes two data augmentation strategies to generate fair augmented graphs and feature representations. Furthermore, we introduce a debiasing learning method that minimizes the dependence between the learned representations and sensitive information to eliminate bias. Extensive experiments on two real-world datasets demonstrate the superiority of our proposed framework. The source code is available at \url{https://github.com/LokHsu/FairDDA}.
\end{abstract}




\maketitle

\section{Introduction}

Recommender systems are widely applied in various domains to help users efficiently discover potential content that they may be interested in \cite{summarize1, social-media, e-commerce}. Numerous algorithms have been proposed to generate high-quality recommendations by analyzing historical interaction patterns. Among them, Graph Convolutional Networks (GCNs) have gained increasing attention for their ability to model complex user preferences \cite{lightgcn,fairgo,chen2024fairgap}. However, in practice, recommender systems often suffer from bias and unfairness, which can raise significant ethical concerns \cite{summarize1,summarize2}. As data-driven applications, recommender systems are susceptible to capturing inappropriate correlations between interaction data and users' sensitive attributes (e.g., gender, age, race, and income level) \cite{li2021user,du2020fairness,pmlr-v81-ekstrand18b}, leading to inherited biases in the representations learned from the input data. To eliminate bias in the learned representations for fairness, many works have been proposed, including regularization-based methods \cite{beyondparity}, adversarial learning-based methods \cite{fairgnn,fairgo}, and disentanglement-based methods \cite{fairmi,fairrec,fairib,faircore}.

Although the above fairness-aware recommendation methods have been shown to be effective in improving fairness through the introduction of specific optimization objectives or constraints during representation learning, they often neglect the detrimental impact caused by the inherent biased in the data.

Specifically, previous studies \cite{burke2018balanced, iosifidis2019fae} have shown that the collected historical data itself may contain inherent unfairness, which can be caused either by imbalanced data distribution or by real-world biases. For example, in the widely used MovieLens dataset, male users contribute more rating behaviors than female users and tend to favor genres such as action and science fiction. This behavioral gap may lead the model to be biased toward the preferences of the male group during training, thereby overlooking female users’ interests in other genres, such as drama or romance \cite{chen2023fda}.

Moreover, GCN-based recommendation models tend to reinforce the \textit{homophily effect} during representation learning \cite{wang2022fairvgnn}, where nodes with similar features are more likely to be connected. As a result, even benign feature dimensions may inadvertently encode sensitive attributes during the graph convolution process. Given that users sharing the same sensitive attributes often exhibit stronger connectivity in social or interaction graphs, the feature propagation mechanism of GCNs may further amplify the impact of these attributes, introducing systemic bias into the recommendation system. Figure 1 illustrates an example of sensitive information propagation in a GCN-based model. In this example, blue and red denote users' sensitive attribute types (e.g., gender), while yellow reflects specific item features and represents user behavioral preferences. On the left side, user and item are connected via interaction relationships, and each user embedding encodes a certain amount of sensitive attributes during preference learning. However, after node embeddings are propagated through the GCN, sensitive attributes further spread across the entire graph structure, causing many user nodes to absorb sensitive attributes from their neighboring nodes during the message-passing process. For example, due to the \textit{homophily effect}, the recommender system makes recommendations to the group of users who interacted with item $v_1$ (i.e., users $u_1$, $u_2$, and $u_3$) based on the blue-colored sensitive attribute rather than their personalized item features. This feature homogenization may mask individual differences among users, leading to unfair outcomes for certain user groups.

\begin{figure}[t]
\centering
\includegraphics[width=.9\linewidth]{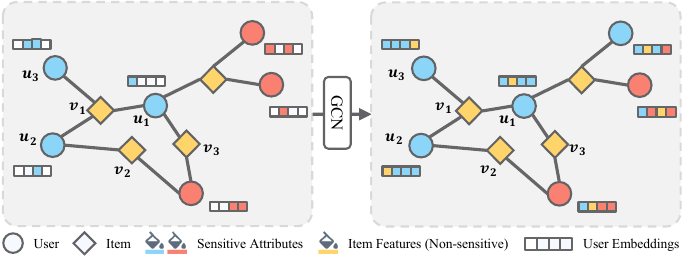}
\vspace{-1em}
\caption{An example of sensitive attributions propagation in a typical graph convolution network.}
\label{fig:1}
\vspace{-1.5em}
\end{figure}

Upon recognizing these issues, some recent studies \cite{agarwal2021nifty,ling2023learning,wang2022fairvgnn,kose2022fairaug,chen2023fda} have focused on addressing unfairness from the perspective of data augmentation, so as to suppress the model's reliance on sensitive attributes by enhancing data diversity and balance. Agarwal et al. \cite{agarwal2021nifty} proposed to construct augmented graphs and feature representations by generating masks with random probabilities. However, randomly masking edges lead to the fragmentation of the original graph structure, limiting the model's ability to capture complete user behavior patterns \cite{simgcl}. Based on the assumption that users from different demographic groups may exhibit similar behaviors, Chen et al. \cite{chen2023fda} generate symmetric interaction data to balance preference distributions across groups. Nevertheless, enforcing symmetry across groups, by generating equivalent interactions for users with similar preferences but different sensitive attributes, will neglect the inherent differences among users from distinct demographic groups.

To ensure the effectiveness of data augmentation while preserving individual users’ behavioral preferences, we propose a \textbf{Fair}ness-aware \textbf{D}ual \textbf{D}ata \textbf{A}ugmentation framework (\textbf{FairDDA}), which combines graph structure augmentation and representation augmentation, aiming to enhance recommendation fairness while preserving utility. Specifically, we first construct fair augmented graphs by pruning sensitive edges (user-item interaction edges that compromise fairness) from the input user-item interaction graph, and generate augmented feature representations by masking sensitive features (features that have a strong correlation with sensitive information) within the input feature representations. Moreover, to improve the effectiveness and fairness of data augmentation, we propose two prior hypotheses to guide the identification of sensitive edges and features. The first assumes that if existing user-item interactions receive lower predicted probabilities in performance-oriented recommendation than in fairness-aware recommendation, the corresponding user-item interaction graph edges are considered as sensitive edges. The second assumes that if a biased representation encoding sensitive information is highly similar to a debiased representation encoding non-sensitive information in the $i$-th feature dimension, then the feature $i$ is considered as a sensitive feature. Based on these hypotheses, we propose two fairness-aware data augmentation strategies to prune sensitive edges and mask sensitive features, thereby generating augmented graph structures and their corresponding representations. Finally, to preserve representation informativeness, we employ contrastive learning to align the representations of the original and augmented graphs, while minimizing their dependence on sensitive information to mitigate potential bias. Our main contributions are listed as follows:

\begin{itemize}[leftmargin=*]
\item We propose a dual data augmentation framework for fairness-aware recommendation, which includes graph structure and feature representation augmentations to effectively remove sensitive information while retaining user preference information.

\item Based on the proposed data augmentation strategies, we design two consistency constraints between the original graph data and its augmented data to preserve utility, and introduce a debiasing learning objective to improve fairness, thereby achieving the optimal balance between fairness and utility in recommendation.

\item We conduct extensive experiments on two real-world datasets. The experimental results show that our method significantly improves recommendation fairness while maintaining utility, outperforming other state-of-the-art fairness-aware methods.
\end{itemize}
\vspace{-1em}

\section{Related Work}
\subsection{Fairness in Recommendation}

The fairness issue in artificial intelligence has recently attracted increasing attention from industry, academia, and society \cite{wang2023survey}. To assess the unfairness of decision models, many fairness concepts have been proposed, with two fundamental ones being individual fairness\cite{individualfairness} and group fairness\cite{groupfairnesssiam}. Group fairness asserts that decisions should be consistent across all demographic groups, while individual fairness emphasizes that similar individuals should receive similar treatment. Recommendation fairness can also be divided into these two categories. In this work, our method focuses on achieving group fairness from the user perspective. To promote group fairness in recommendation, researchers have proposed several approaches. Yao and Huang \cite{beyondparity} introduced four statistical fairness constraints to optimize recommendation fairness. Some previous works used adversarial networks to optimize feature representations to obscure sensitive information \cite{adv,fairgo,fairgnn}. Some recent studies have applied the concept of disentanglement to fair representation learning, encoding information representations of different factors and dissociating these factors through dependence measures methods \cite{fairrec,fairmi,sarhan2020orthogonal,guo2022distance,oh2022distributional,fairib,faircore}. For example, Xie et al. \cite{fairib} and Zhao et al. \cite{fairmi} aimed to minimize the MI between these two representations. Bin et al. \cite{faircore} further introduce an attribute-neutralization strategy based on MI optimization to learn attribute-invariant fair representations, aiming to improve recommendation fairness. However, due to the different learning objectives of fairness-aware recommendation and the performance-oriented recommendation, it is hard for the existing fairness recommendation methods to ensure both utility and fairness.

\subsection{Data Augmentation in Fairness}

Data augmentation techniques were initially developed to generate diverse data, enabling models to learn richer features and improve performance on unseen samples. Researchers have investigated data augmentation methods for fairness in machine learning, which shows promising potential. Chen et al. \cite{fairfil} proposed a text augmentation technique that replaces sensitive attribute words in sentences, improving fairness for pre-trained sentence encoders. D'Inc{\`a} et al. \cite{image_da} applied feature editing to input images, generating augmented images to mitigate bias. In natural language processing and computer vision tasks, input data such as text and images exhibits inherent structural and symmetric properties that allow data augmentation through semantic or geometric transformations\cite{shorten2021da_text,shorten2019da_image}. In contrast, the data used in recommendation tasks is less structured, and characterized by complex and nonlinear user behavior, which is hard to directly apply traditional augmentation strategies. Thus, methods such as graph-based augmentation and feature-level augmentation for fairness have been proposed \cite{chen2023fda,ling2023learning,ma2022graphcounterfactual}. Recently, Chen et al. \cite{chen2023fda} propose a data augmentation method for recommendation fairness based on the assumption that users from different groups have similar preferences. However, this method neglects the inherent unique preferences of each user group and introduces substantial distortion to user preferences when generating synthetic behavioral data by adding random noise, resulting in significantly decreased utility and limited fairness improvement.

In this work, ensuring that the augmented data is fair and retains information comparable to the original graph, we first introduce two prior hypotheses to explicitly define sensitive information in graph data. Building on this, our model can accurately mitigate sensitive information during data augmentation while preserving user preferences, thereby achieving fair and precise recommendations.

\section{Preliminary}
\subsection{Problem Formulation}

Let $\mathcal{U}=\{u_1,\ldots,u_M\}$ and $\mathcal{V}=\{v_1,\ldots,v_N\}$ denote user and item sets, with sizes $M$ and $N$ respectively. The sensitive attribute of user $u$ is represented as a one-hot vector $s_u=[s_{u,c}]_{c=0}^{C-1}$ in $C$ dimensions, where $C$ is the number of classes (e.g., $C=2$ for gender). User-item interactions form an adjacency matrix $\mathbf{A} \in \mathbb{R}^{M \times N}$, where $\mathbf{A}_{uv}=1$ if user $u$ interacted with item $v$, and $\mathbf{A}_{uv}=0$ otherwise. For each user $u$, $R_u=\{v| \mathbf{A}_{uv} =1\}$ denotes interacted items. And users, items and their interactions naturally form an interaction graph $\mathcal{G}=\{\mathcal{U},\mathcal{V},\mathcal{E}\}$, where $\mathcal{E}=\{(u,v)|u\in \mathcal{U},v \in \mathcal{V},\mathbf{A}_{uv}=1\}$ denotes the edge set. In this work, we aim to design a recommendation model that achieves both high-quality recommendations and fairness. If users from different demographic groups receive systematically different recommendations despite similar preferences, the system violates fairness principles. A fair model should ensure balanced exposure and equal relevance across groups. To this end, we adopt two metrics to measure fairness: \textit{Demographic Parity} \cite{dwork2012dp}, which evaluates the balance of recommendation distribution across different groups; and \textit{Equal Opportunity} \cite{hardt2016eo}, which measures whether users with similar interests receive equal recommendation opportunities regardless of their group membership. The formal definitions are provided in Section~\ref{sec:metrics}.

\subsection{GCNs in Recommendation}

To improve recommendation performance, modern recommendation models typically employ GCNs to enhance these initial representations. Given a user-item interaction graph $\mathcal{G}=\{\mathcal{U},\mathcal{V},\mathcal{E}\}$, the feature representations of users/items can be iteratively updated through multiple GCN layers as:
\begin{equation}
    \mathbf{X}_U^{(l+1)} = \phi\left( \hat{\mathbf{A}} \mathbf{X}_V^{(l)} \mathbf{W}_U^{(l)}\right), \quad \mathbf{X}_V^{(l+1)} = \phi\left( \hat{\mathbf{A}}^\top \mathbf{X}_U^{(l)} \mathbf{W}_V^{(l)}\right),
\end{equation}
where $\hat{\mathbf{A}} = \mathbf{D}_U^{-1/2} \mathbf{A} \mathbf{D}_V^{-1/2}$; $\mathbf{D}_U$ and $\mathbf{D}_V$ are the degree matrices representing the degrees of users and items, respectively; $\mathbf{W}_U^{(l)}$ and $\mathbf{W}_V^{(l)}$ are the learnable weight matrices applied to the user and item sides, respectively; and $\phi$ denotes a nonlinear activation function. $\mathbf{X}_U^{(l+1)}$ and $\mathbf{X}_V^{(l+1)}$ represent the output embeddings of users and items at the $(l+1)$-th GCN layer, respectively. At the input layer, $\mathbf{X}_U^{(0)}$ and $\mathbf{X}_V^{(0)}$ are randomly initialized feature representations.

Among various GCN variants, LightGCN \cite{lightgcn} stands out by discarding feature transformation and nonlinear activation, making it both lightweight and effective for recommendation tasks. Its propagation rule is defined as follows:
\begin{equation}
     \mathbf{X}_U^{(l+1)} =  \boldsymbol{\hat{A}} \mathbf{X}_V^{(l)},\quad
     \mathbf{X}_V^{(l+1)} =  \boldsymbol{\hat{A}}^\top \mathbf{X}_U^{(l)}.
\end{equation}

The final user and item representations are obtained by averaging the outputs from all $L$ GCN layers. Formally, the final embeddings can be expressed as follows:
\begin{equation}
    \mathbf{X}_U=\frac{1}{L+1} \sum_{l=0}^L \mathbf{X}_U^{(l)},\quad
    \mathbf{X}_V=\frac{1}{L+1} \sum_{l=0}^L \mathbf{X}_V^{(l)}.
\end{equation}

To ensure comparable performance, in this work, we follow existing SOTA baselines and adopt LightGCN as the base encoder.

\section{Methodology}

\begin{figure*}[t]
\centering
\includegraphics[width=.94\linewidth]{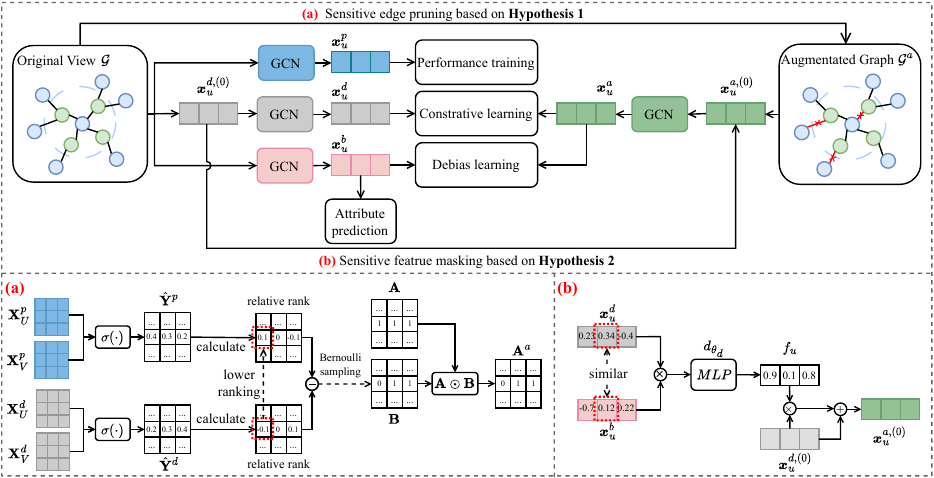}
\vspace{-.2em}
\caption{The framework of the proposed method, with details shown only for the user side. The key differences from the item-side structure are as follows: i) The biased item representations are trained without the attribute prediction task, and ii) the debiasing learning method is not applied to the item side.}
\label{fig:framework}
\vspace{-.5em}
\end{figure*}

\subsection{Overview}

The proposed method is illustrated in Figure~\ref{fig:framework}. Our goal is to establish a fairer data distribution to train an effective and unbiased recommendation model. Specifically, given the input graph $\mathcal{G}=\{\mathcal{U},\mathcal{V},\mathcal{E}\}$, we first construct three distinct feature representations for users and items, each serving different purposes: i) The performance-oriented representations $\mathbf{X}^p=[\mathbf{X}_U^p;\mathbf{X}_V^p]$, which encode rich preference information while inevitably mixing some sensitive attributes. ii) The biased representations $\mathbf{X}^b=[\mathbf{X}_U^b;\mathbf{X}_V^b]$, which capture rich sensitive information. iii) The debiased representations $\mathbf{X}^d=[\mathbf{X}_U^d;\mathbf{X}_V^d]$, which preserve user preference information while eliminating sensitive information for fair recommendation. Next, based on the input graph and the debiased representations, we construct an augmented graph $\mathcal{G}^a=\{\mathcal{U},\mathcal{V},\mathcal{E}^a\}$, and its corresponding augmented representations $\mathbf{X}^a=[\mathbf{X}_U^a;\mathbf{X}_V^a]$ using two data augmentation strategies: sensitive edge pruning and sensitive feature masking. To alleviate potential information loss in the augmented representations and enhance the quality of debiased representations, we encourage the consistency between the original graph data and its augmented data through graph contrastive learning. Finally, we introduce a debiasing learning method to optimize the fairness of the augmented graph and its representations.

\subsection{Hypothesis of Sensitive Information}

Striking a proper balance between fairness and utility in recommendation requires carefully pruning sensitive edges and masking sensitive features, while minimizing disruption to user preferences. Therefore, we first propose two prior hypoheses to provide clear definitions for sensitive edges and sensitive features:

\noindent \underline{\textbf{Hypothesis 1}}. Suppose there are performance-oriented representations $\boldsymbol{x}_u^p$ and $\boldsymbol{x}_v^p$, and debiased representations $\boldsymbol{x}_u^{d}$ and $\boldsymbol{x}_v^{d}$. Let $\hat Y_u^{d}$ be the predicted recommendations for the user $u$ based on $\boldsymbol{x}_u^{d}$ and $\boldsymbol{x}_v^{d}$ , and $\hat Y_u^p$ be the recommendations based on $\boldsymbol{x}_u^p$ and $\boldsymbol{x}_v^p$. $\hat Y_u^{d}$ is considered fairer than $\hat Y_u^p$. If item $v$ is ranked lower in $\hat Y_u^{d}$ than in $\hat Y_u^p$, it indicates that the interaction between $u$ and $v$ forms a sensitive edge. Down-weighting the importance of such an edge in the graph is expected to help improving recommendation fairness.

\noindent \underline{\textbf{Hypothesis 2}}. Suppose there are a debiased representation $\boldsymbol{x}^d$ and a biased representation $\boldsymbol{x}^{b}$ for a given user or item. The former encodes user preference information while excluding sensitive information, whereas the latter encodes rich sensitive information. If the debiased representation $\boldsymbol{x}^d$ is highly similar to the biased representation $\boldsymbol{x}^b$ at a given feature dimension $i$, then feature $i$ is considered a sensitive feature. Avoiding such a feature's similarity is expected to promote a fairer representation.

\subsection{Representation generation}
\subsubsection{\hspace{-0.1em}\textbf{Performance-oriented Representation}}

\hspace{-0.4em} The performance-oriented representations $\mathbf{X}^p$ are designed to encode user preference information from interaction data. To generate high-quality representations for subsequent fine-tuning, we train a performance-oriented model (e.g., LightGCN) using BPR loss \cite{rendle2012bpr}:
\begin{equation}
\label{eq:4}
    \min_{\mathbf{X}^{p,(0)}} \mathcal{L}_{BPR}= - \frac{1}{|\mathcal{D}|}\sum_{(u,v,j) \in \mathcal{D}} \ln \sigma(\boldsymbol{x}_u^p \cdot \boldsymbol{x}_{v}^p-\boldsymbol{x}_u^p \cdot \boldsymbol{x}_{j}^p),
\end{equation}
where $\sigma(x)=\frac{1}{1+e^{-x}}$ is the sigmoid function, and $\mathcal{D}$ is the training set consisting of triplets $(u, v, j)$. In each triplet, $v$ is a positive item interacted by user $u$, and $j$ is a negative item not interacted by user $u$. Here, $\boldsymbol{x}_u^p$, $\boldsymbol{x}_{v}^p$ and $\boldsymbol{x}_j^p$ are the performance-oriented representations within $\mathbf{X}^p$ for the user $u$ and items $v$ and $j$, respectively. Parameters of the performance-oriented model are saved upon reaching optimal utility performance, serving as a pre-trained model.

\subsubsection{\textbf{Biased Representation}}

The biased representations are intended to capture and encode sensitive information. To achieve this, the biased representations are trained through an attribution prediction task, where the higher the prediction accuracy indicates the more sensitive information is encoded. In particular, to obtain the predicted attributes, the biased representations of users are fed into an attribute classifier $c_{\theta_c}$, which is implemented by a one-layer fully-connected neural network with parameters $\theta_c$. Given a biased representation $\boldsymbol{x}_u^b \in \mathbf{X}_U^b$, the attribute prediction process can be formulated as:
\begin{equation}
    \hat s_u=\text{softmax}(c_{\theta_c}(\boldsymbol{x}_u^b)),
\end{equation}
where $\hat s_u$ is the sensitive attribute prediction of user $u$. To improve the prediction accuracy with the goal of capturing rich sensitive information, we construct the cross-entropy loss to optimize the bias representations as follows:
\begin{equation}
\label{eq:6}
    \min_{\mathbf{X}^{b,(0)}, \theta_c} \mathcal{L}_{CE}=  -\frac{1}{M} \sum_{u=1}^{M} \sum_{c=1}^C (s_{u,c} \cdot \log(\hat s_{u,c})),
\end{equation}
where $s_u=[s_{u,c}]_{c=0}^{C-1}$ is the sensitive attribute of the user $u$.

\subsubsection{\textbf{Debiased representation}}

The debiased representations $\mathbf{X}^d$, generated by refining both the performance-oriented and biased representations, are designed to mitigate the negative impact of the imbalanced graph $\mathcal{G}$ and used for the final recommendation. Specifically, to learn the debiased representations, we first propose two data augmentation strategies to construct a fair augmented graph $\mathcal{G}^a$ and its corresponding augmented representation $\mathbf{X}^a$. Building on this, we introduce a debiasing learning approach that minimizes the dependency between the augmented representations $\mathbf{X}^a$ and the biased representations $\mathbf{X}^b$ (which encode sensitive information), in order to eliminate the bias in the augmented graph data and enhance fairness. Finally, we propose a graph contrastive learning method to transfer semantic information from the augmented graphs into the debiased representations $\mathbf{X}^d$.

A detailed description of the two data augmentation strategies and the above representation learning process is presented below.

\subsection{Fair Data Augmentation}
\subsubsection{\textbf{Sensitive Edge Pruning}}
\label{sec:4.4.1}

Given a graph $\mathcal{G}=\{\mathcal{U},\mathcal{V},\mathcal{E}\}$ with the corresponding adjacency matrix $\mathbf{A}$, our method constructs a fair augmented graph $\mathcal{G}^a=\{\mathcal{U},\mathcal{V},\mathcal{E}^a\}$ by pruning sensitive edges, which can be formulated as:
\begin{equation}
    \mathbf{A}^a= \mathbf{A} \odot \mathbf{B},
\end{equation}
where $\mathbf{A}^a$ is the adjacency matrix of the augmented graph $\mathcal{G}^a$, and its corresponding edge set is defined as: $\mathcal{E}^a=\{(u,v)|(u,v)\in \mathcal{E}, \mathbf{A}_{uv}^a=\mathbf{A}_{uv}\cdot\mathbf{B}_{uv}\}$, where $\mathbf{B} \in {\{0,1\}}^{M \times N}$ is a binary edge pruning matrix sampled from a Bernoulli distribution \cite{statistical}. Given an element $ \mathbf{B}_{uv} \in \mathbf{B}$, there is $\mathbf{B}_{uv} \sim \mathcal{B}(p_{uv})$, where $p_{uv}$ denotes the probability that the edge $(u,v)$ is pruned from $\mathcal{G}$. Therefore, the key challenge of pruning sensitive edge lies in how to determine the pruning probability $p_{uv}$ for each edge. According to \underline{\textbf{Hypothesis 1}}, we aim to derive the pruning probabilities based on the ranking changes of items across different goal-oriented recommendations, as these changes reflect the impact of user interactions on fairness.

Specifically, we first derive the performance-oriented recommendation $\hat{Y}_u^p$ using representations $\mathbf{X}^p$. For a user $u$ and an interacted item $v \in R_u$, the predicted score $\hat{y}_{uv}^p \in \hat{Y}_u^p$ is given by:
\begin{equation}
    \hat{y}_{uv}^{p}=\sigma(\boldsymbol{x}_u^p \cdot \boldsymbol{x}_{v}^p).
\end{equation}
To make items comparable across different recommendation rankings, we calculate the relative ranking of item $v$ in the performance-oriented recommendation based on the average predicted scores of all interacted items in $R_u$, which can be expressed as:
\begin{equation}
     \Delta r^p_u(v)=\hat{y}_{uv}^{p}- \bar y_u^p,\quad
     \bar y_u^p=\frac{1}{|R_u|}\sum_{j \in R_u}\hat y_{uj}^{p},
\end{equation}
where $R_u$ is the set of all items interacted by user $u$. Based on the debiased representations $\mathbf{X}^d$, we compute the fairness-oriented recommendation $\hat{Y}_u^d$ and derive the relative ranking $\Delta r_u^d(v)$ in the same way. Then, the ranking difference of item $v$ between the two recommendations is used to calculate the pruning probability $p_{uv}$ for edge $(u,v)$:
\begin{equation}
    p_{uv}=\exp(\Delta r^d_u(v)-\Delta r^p_u(v)).
\end{equation}

Since probabilities must be within the range $[0, 1]$, we clip $p_{uv}$ to ensure it falls within this interval by:
\begin{equation}
    p_{uv}=\begin{cases}
        \;1,& \text{if}~p_{uv}>1 , \\
        \;p_{uv},& \text{otherwise}.
    \end{cases}
\end{equation}

As a result of these operations, we obtain an edge pruning indicator $\mathbf{B}_{uv} \in \mathbf{B}$ according to the Bernoulli sampling with probability $p_{uv}$, i.e., $\mathbf{B}_{uv} \sim \mathcal{B}(p_{uv})$. This process will be repeated for all users to produce the complete edge pruning matrix $\mathbf{B}$. Additionally, we note that the matrix $\mathbf{B}$ is non-differentiable, which prevents gradient-based optimization during end-to-end training. To address this issue, we approximate the Gumbel softmax trick \cite{Jang2017gumbel} is used to approximate Bernoulli sampling. Given a probability $p_{uv}$ for user $u$ and item $v$, the continuous approximation is computed as:
\begin{equation}
    \hat B_{uv}=\frac{1}{1+\exp(-(\log(p_{uv})+g)/\tau)},
\end{equation}
where $\tau$ is a temperature factor and $g \sim \text{Gumbel}(0,1)$ represents random noise. For forward propagation, the discrete value $B_{uv} = \lfloor \hat{B}_{uv} + 1/2 \rfloor$ is used as the output of the Bernoulli sampling. For backward propagation, the gradient is approximated as $\Delta_{\psi} \hat{B}_{uv} ~\approx ~ \Delta_{\psi}B_{uv}$.

\subsubsection{\textbf{Sensitive Feature Masking}}

Following \underline{\textbf{Hypothesis 2}}, we propose a sensitive feature masking strategy to generate fair augmented representations by masking the sensitive features of the input representations. Specifically, given initial debiased representations $\mathbf{X}_U^{d,(0)}$ and $\mathbf{X}_V^{d,(0)}$, we generate the initial augmented representations $\mathbf{X}_U^{a,(0)}$ and $\mathbf{X}_V^{a,(0)}$ as follows:
\begin{equation}
\begin{split}
    \mathbf{X}_U^{a,(0)} &= \mathbf{X}_U^{d,(0)} + \mathbf{X}_U^{d,(0)} \odot \mathbf{F}_U,\\
    \mathbf{X}_V^{a,(0)} &= \mathbf{X}_V^{d,(0)} + \mathbf{X}_V^{d,(0)} \odot \mathbf{F}_V,
\end{split}
\end{equation}
where $\mathbf{F}_U \in \mathbb{R}^{M\times d}$ and $\mathbf{F}_V \in \mathbb{R}^{N \times d}$ are feature mask matrices generated by a sensitive feature detector $d_{\theta_d}$, which is implemented by a multi-layer neural network. In particular, to construct these feature mask matrices, we calculate individual feature mask vectors for each user and item based on their debiased and biased representations. Specifically, Given the debiased representation $\boldsymbol{x}_u^d$ and $\boldsymbol{x}_v^d$, and the biased representation $\boldsymbol{x}_u^b$ and $\boldsymbol{x}_v^b$, we compute the feature mask vectors $\boldsymbol{f}_u \in \mathbf{F}_U$ and $\boldsymbol{f}_v \in \mathbf{F}_V$ as follows:
\begin{equation}
\label{eq:feature-1}
\begin{split}
    \boldsymbol{f}_u&=\exp(-\sigma(d_{\theta_d}(\boldsymbol{x}_u^d \odot \boldsymbol{x}_u^b))),\\
    \boldsymbol{f}_v&=\exp(-\sigma(d_{\theta_d}(\boldsymbol{x}_v^d \odot \boldsymbol{x}_v^b))).
\end{split}
\end{equation}

\subsection{Debiasing Learning}

Since the biased representation encodes users' sensitive information, to improve the model's fairness, our method aims to exclude the sensitive information from the augmented representations. To achieve this, we resort to minimizing the dependence between augmented representations $\mathbf{X}_U^a$ and biased representations $\mathbf{X}_U^b$ using the Hilbert-Schmidt Independence Criterion (HSIC) \cite{gretton2005HSIC}, which is a measurement of the dependence between random variables. Given $m$ i.i.d sample pairs ${(x_u^a, x_u^b)}_{u=1}^m$, derived from $\mathbf{X}_U^a$ and $\mathbf{X}_U^b$, where $m$ is the number of user-item interactions retained after the sensitive edge pruning (Section \ref{sec:4.4.1}), and following prior studies \cite{ma2020hsic,fairib}, the HSIC is computed as follows:
\begin{equation}
    \text{HSIC}(\mathbf{X}_U^a, \mathbf{X}_U^b) = \frac{ \text{trace}(\mathbf{K}_a\mathbf{H}\mathbf{K}_b\mathbf{H})}{(m-1)^2},
\end{equation}
where $\mathbf{H}$ is the centering matrix defined as $\mathbf{H}=\mathbf{I}-\frac{1}{m}\boldsymbol{1}\boldsymbol{1}^\top$. Here, $\mathbf{I}$ is the identity matrix of size $m \times m$, and $\boldsymbol{1}$ is a column vector of ones with dimension $m$. $\mathbf{K}_a \in \mathbb{R}^{m \times m}$ and $\mathbf{K}_b \in \mathbb{R}^{m \times m}$ are kernel matrices of $\mathbf{X}_U^a$ and $\mathbf{X}_U^b$, where $\mathbf{K}_{ij} \in \mathbf{K}_a$ is calculated by the kernel function $k(x_i,x_j)$. We use the Radial Basis Function (RBF) \cite{ham2004kernel} as the kernel function, which is formulated as:
\begin{equation}
    k(x_i,x_j) = \exp\left(-\frac{||x_i - x_j||^2}{2\sigma^2}\right).
\end{equation}

Generally, $\text{HSIC}(\mathbf{X},\mathbf{Y}) \geq 0$, particularly $\text{HSIC}(\mathbf{X},\mathbf{Y}) = 0$ if and only if variables $\mathbf{X}$ and $\mathbf{Y}$ are independent. To make the augmented representations $\mathbf{X}_U^a$ independent of sensitive information, we introduce an HSIC-based debiasing objective, as follows:
\begin{equation}
    \min_{\mathbf{X}^{d,(0)},{\theta_d}} \mathcal{L}_{dl}= \text{HSIC}(\mathbf{X}_U^a,\mathbf{X}_U^b).
\end{equation}

\subsection{Graph Contrastive Learning}
\label{sec:ct}

To avoid damaging the structural integrity of the interaction graph and potentially losing critical information due to edge removal in the augmented graph, we design the following graph reconstruction loss:
\begin{equation}
    \min_{\mathbf{X}^{d,(0)},{\theta_d}} \mathcal{L}_{recon}=- \frac{1}{\mathcal{|\mathcal{D}|}}\sum_{(u,v,j) \in \mathcal{D}} \ln \sigma(\boldsymbol{x}_u^a \cdot \boldsymbol{x}_{v}^a-\boldsymbol{x}_u^a \cdot \boldsymbol{x}_{j}^a),
\end{equation}
where $\mathcal{D}$ denotes the training set of the original interaction graph, while $\boldsymbol{x}_u^a$ and $\boldsymbol{x}_v^a$ indicate the user and item representations generated from the augmented graph, respectively. By preserving the relative preference of positive items over negative ones, this objective ensures that the embeddings of the augmented graph retain comparable recommendation quality to those of the original graph and prevent excessive edge removal.

Moreover, to maintain the informativeness of augmented representations, we define a contrastive learning objective to optimize the similarity between the debiased representation and its augmented representation. The contrastive loss is computed from the debiased embedding $\boldsymbol{x}_u^d\in\mathbf{X}_U^{d}$ and the augmented embedding $\boldsymbol{x}_u^a\in\mathbf{X}_U^{a}$:
\begin{equation}
\begin{split}
    &\mathcal{L}_{cl}(\boldsymbol{x}_u^d,\boldsymbol{x}_u^a)\\
    &=-\log \frac{\exp(\text{sim}(\boldsymbol{x}_u^d,\boldsymbol{x}_u^a))}{\frac{1}{M}\sum_{u'}^M [\exp(\text{sim}(\boldsymbol{x}_u^d,\boldsymbol{x}_{u'}^a))+ \exp(\text{sim}(\boldsymbol{x}_u^d,\boldsymbol{x}_{u'}^d))]},
\end{split}
\end{equation}
where $\text{sim}(\cdot)$ is implemented with the cosine similarity function. Therefore, the overall contrastive learning loss is formulated from both the user and item sides as follows:
\begin{equation}
    \begin{split}
        \min_{\mathbf{X}^{d,(0)},{\theta_d}}\ \mathcal{L}_{cl} = 
        & \frac{1}{M} \sum_{u=1}^M \left[\mathcal{L}_{cl}(\boldsymbol{x}_u^d,\boldsymbol{x}_u^a)+\mathcal{L}_{cl}(\boldsymbol{x}_u^a,\boldsymbol{x}_u^d)\right] \\
        & + \frac{1}{N} \sum_{v=1}^N \left[\mathcal{L}_{cl}(\boldsymbol{x}_v^d,\boldsymbol{x}_v^a)+\mathcal{L}_{cl}(\boldsymbol{x}_v^a,\boldsymbol{x}_v^d)\right].
    \end{split}
\end{equation}

\subsection{Model Training}
\subsubsection{\textbf{Training Phases}}

The overall training consists of a pre-training phase for learning both performance-oriented and biased representations via Equations (\ref{eq:4}) and (\ref{eq:6}), followed by a main training phase where the BPR loss is adopted as the main loss function:
\begin{equation}
\mathcal{L}_{BPR}=- \frac{1}{\mathcal{|D|}}\sum_{(u,v,j) \in \mathcal{D}} \ln \sigma(\boldsymbol{x}_u^d \cdot \boldsymbol{x}_{v}^d-\boldsymbol{x}_u^d \cdot \boldsymbol{x}_{j}^d).
\end{equation}
To optimize the debiased representations for fair recommendation, we incorporate auxiliary loss terms into the main loss, resulting in the overall optimization objective for the main training phase:
\begin{equation}
\min_{\mathbf{X}^{d,(0)},{\theta_d}} \mathcal{L}_{all}= \mathcal{L}_{BPR} + \lambda_r\mathcal{L}_{recon} + \lambda_c\mathcal{L}_{cl} +\lambda_d\mathcal{L}_{dl},
\end{equation}
where $\lambda_r$, $\lambda_c$ and $\lambda_d$ are hyperparameters.

\subsubsection{\textbf{Time Complexity Analysis}}

The overall time complexity of FairDDA mainly stems from graph convolution operations and the loss computations involved in debiasing and graph contrastive learning. The detailed time complexity for each training epoch is analyzed as follows: The LightGCN backbone is employed in both the pre-training and main training phases, each incurring a computational cost of $\mathcal{O}(L(M + N)d)$, where $d$ is embedding dimension, and $(M + N)$ denotes the total number of users and items. Moreover, during the main training phase, the debiasing learning constructs $m \times m$ kernel matrices, applies centering and matrix multiplications, resulting in an additional computational overhead of $\mathcal{O}(m^2d + m^3)$. The graph contrastive learning performs global pairwise similarity computation and normalization over all $M$ users and $N$ items, introducing a complexity of $\mathcal{O}((M+N)^2 d)$. Furthermore, although the proposed sensitive edge pruning-based augmented graph construction incurs some additional computational overhead compared to existing fair recommendation methods \cite{fairmi,fairib}, its complexity is linear in the cost of graph convolution, i.e., $\mathcal{O}((M + N)d)$, and thus remains computationally efficient.

\section{Experiments}

In this section, we conduct extensive experiments to evaluate the performance of FairDDA and address the following questions:

\begin{itemize}[leftmargin=*]
\item \textbf{RQ1}: How effective is FairDDA for fair recommendation compared to its competitors?

\item \textbf{RQ2}: How does each module of FairDDA affect its recommendation performance?

\item \textbf{RQ3}: How do loss coefficient settings influence the trade-off between fairness and utility in recommendation models?

\item \textbf{RQ4}: Can FairDDA better balance fairness and preference in user embeddings compared to the baseline?
\end{itemize}

\vspace{-.4em}
\subsection{Experimental Setup}
\subsubsection{\textbf{Datasets}}

We conduct experiments on two commonly used datasets: Movielens-1M (abbreviated as ML-1M) \cite{movielens} and LastFM-360K (abbreviated as LastFM) \cite{LastFM}. The statistics of each dataset are detailed in Table~\ref{tab:dataset}. Besides the basic user-item interaction data, both datasets also contain user gender information. To ensure comparability and reproducibility, we adopt the same experimental settings as prior work \cite{fairmi, fairib, faircore}. Due to space limitations, this study focuses on evaluating \textbf{gender fairness} (binary attribute scenario) to verify the effectiveness of our method. See Appendix \ref{sec:multi-sensitive} for experiments on multi-class sensitive attributes.


\begin{table}[t]
    \centering
    \caption{Description of the dataset. \label{tab:dataset}}
    \vspace{-.7em}
    \begin{tabular}{cccccc}
    \hline
    \textbf{Dataset} & \textbf{\#Users} & \textbf{\#Items} & \textbf{\#Iteractions} & \textbf{Density} \\
    \hline
     Movielens-1M & 6,040 & 3,952 & 1,000,209 & 4.19\% \\
     LastFM-360K & 48,386 & 21,711 & 2,045,305 & 0.19\% \\
    \hline
    \end{tabular}
    \vspace{-.9em}
\end{table}

\vspace{-.1em}
\subsubsection{\textbf{Implement Details}}

We implement our method in PyTorch 2.1.2 using the Adam optimizer with a learning rate of 0.001. The regularization parameter is set to 0.001. All compared methods use a representation vector dimension $d$ of 64, with batch sizes of 2048 for ML-1M and 4096 for LastFM. The values of $\lambda_r$, $\lambda_c$, and $\lambda_d$ are selected via grid search over $\{0.1, 0.5, 1, 5, 10\}$, $\{0.01, 0.05, 0.1, 0.5, 1\}$, and $\{10, 20, 30, 40, 50\}$, respectively. For the LightGCN backbone, the number of propagation layers $L$ is set to 3. We report the average performance over 10 runs to ensure reliability.

\begin{table*}[!t]
    \caption{Performance comparison of recommendation methods on utility and fairness metrics across two datasets. The upward arrow ($\uparrow$) indicates higher values are better, while the downward arrow ($\downarrow$) indicates lower values are preferable.}
    \label{tab:performance}
    \centering
    \resizebox{\linewidth}{!}{
    \begin{tabular}{p{1.1cm}|c|ccc|ccc|ccc|ccc}
        \hline  
        \centering \multirow{2}{*}{\textbf{Dataset}} & \multirow{2}{*}{\textbf{Method}} & \multicolumn{3}{c|}{ \textit{\textbf{NDCG@K}} $\uparrow$} & \multicolumn{3}{c|}{ \textit{\textbf{Recall@K}} $\uparrow$} & \multicolumn{3}{c|}{\textit{\textbf{DP@K}} $\downarrow$} & \multicolumn{3}{c}{\textit{\textbf{EO@K}} $\downarrow$} \\
        & & \textbf{10} & \textbf{20} & \textbf{30} & \textbf{10} & \textbf{20} & \textbf{30} & \textbf{10} & \textbf{20} & \textbf{30} & \textbf{10} & \textbf{20} & \textbf{30} \\
        \hline
        \multirow{12}{*}{\textbf{ML-1M}}   
        & \textbf{LightGCN} & 0.2018 & 0.2671 & 0.3060 & 0.1511 & 0.2449 & 0.3085 & 0.2919 & 0.2626 & 0.2449 & 0.3609 & 0.3325 & 0.3085 \\
        \cline{2-14}
        & \textbf{BP} & 0.1961 & 0.2603 & 0.2938 & 0.1422 & 0.2363 & 0.2828 & 0.2097 & 0.1924 & 0.1545 & 0.3047 & 0.2955 & 0.2611 \\
        & \textbf{Adv} & 0.1963 & 0.2579 & 0.2975 & 0.1469 & 0.2346 & 0.2998 & 0.1532 & 0.1183 & 0.1068 & 0.2694 & 0.2338 & 0.2203 \\
        & \textbf{FDA} & 0.1873 & 0.2539 & 0.2878 & 0.1419 & 0.2316 & 0.2940 & 0.2637 & 0.1889 & 0.2174 & 0.3443 & 0.2986 & 0.2858 \\
        & \textbf{FairRec} & 0.1950 & 0.2561 & 0.2955 & 0.1472 & 0.2339 & 0.2986 & 0.1536 & 0.1193 & 0.1042 & 0.2590 & 0.2283 & 0.2243 \\
        & \textbf{FairGo} & 0.1822 & 0.2373 & 0.2741 & 0.1336 & 0.2108 & 0.2710 & 0.2728 & 0.2436 & 0.2275 & 0.3382 & 0.3101 & 0.2921 \\
        & \textbf{FairGNN} & 0.1964 & 0.2569 & 0.2963 & 0.1466 & 0.2323 & 0.2969 & 0.1472 & 0.1181 & 0.1045 & 0.2608 & 0.2320 & 0.2221 \\
        & \textbf{FairMI} & 0.2128 & \underline{0.2752} & 0.3148 & 0.1586 & \underline{0.2477} & \underline{0.3124} & 0.1337 & 0.1111 & 0.1004 & 0.2228 & 0.2006 & 0.1979 \\
        & \textbf{FairIB} & 0.2001 & 0.2655 & 0.3003 & 0.1505 & 0.2401  & 0.3063 & 0.1405 & 0.1137  & 0.1035 & 0.2038 & 0.1781 & \underline{0.1688} \\
        & \textbf{FairCoRe} & \underline{0.2136} & 0.2751 & \underline{0.3150} & \underline{0.1591} & 0.2462 & 0.3116 & \underline{0.1225} & \underline{0.0999} & \underline{0.0897} & \underline{0.1983} & \underline{0.1591} & 0.1755 \\
        \cline{2-14}
        & \textbf{FairDDA} & \textbf{0.2138} & \textbf{0.2763} & \textbf{0.3164} & \textbf{0.1601} & \textbf{0.2487} & \textbf{0.3139} & \textbf{0.1183} & \textbf{0.0961} & \textbf{0.0863} & \textbf{0.1833} & \textbf{0.1388} & \textbf{0.1575} \\
        & $\mathbf{p}$\textbf{-value} & 2.1$\mathrm{e}{-2}$ & 2.5$\mathrm{e}{-3}$ & 1.2$\mathrm{e}{-3}$ & 3.4$\mathrm{e}{-2}$ & 6.1$\mathrm{e}{-2}$ & 2.3$\mathrm{e}{-2}$ & 4.7$\mathrm{e}{-3}$ & 3.5$\mathrm{e}{-3}$ & 1.1$\mathrm{e}{-2}$ & 6.0$\mathrm{e}{-3}$ & 7.4$\mathrm{e}{-4}$& 9.8$\mathrm{e}{-4}$ \\
        \hline
        \hline
        \multirow{12}{*}{\textbf{LastFM}}
        & \textbf{LightGCN} & 0.1971 & 0.2463 & 0.2762 & 0.1572 & 0.2381 & \textbf{0.2964} & 0.2860 & 0.2673 & 0.2569 & 0.3508 & 0.3332 & 0.3247 \\
        \cline{2-14}
        & \textbf{BP} & 0.1863 & 0.2403 & 0.2619 & 0.1484 & 0.2327 & 0.2865 & 0.2201 & 0.2037 & 0.1965 & 0.3012 & 0.3008 & 0.2867 \\
        & \textbf{Adv} & 0.1887 & 0.2384 & 0.2652 & 0.1499 & 0.2312 & 0.2853 & 0.1382 & 0.1248 & 0.1163 & 0.2682 & 0.2617 & 0.2576 \\
        & \textbf{FDA} & 0.1703 & 0.2389 & 0.2421 & 0.1362 & 0.2278 & 0.2787 & 0.2294 & 0.1946 & 0.2062 & 0.3017 & 0.2712 & 0.2829 \\
        & \textbf{FairRec} & 0.1892 & 0.2375 & 0.2627 & 0.1505 & 0.2276 & 0.2837 & 0.1397 & 0.1295 & 0.1157 & 0.2700 & 0.2607 & 0.2596 \\
        & \textbf{FairGo} & 0.1693 & 0.2115 & 0.2389 & 0.1371 & 0.2065 & 0.2627 & 0.2626 & 0.2450 & 0.2338 & 0.3282 & 0.3124 & 0.3059 \\
        & \textbf{FairGNN} & 0.1879 & 0.2358 & 0.2651 & 0.1501 & 0.2290 & 0.2859 & 0.1372 & 0.1210 & 0.1171 & 0.2690 & 0.2609 & 0.2592 \\
        & \textbf{FairMI} & 0.1976 & 0.2464 & 0.2766 & \textbf{0.1576} & \underline{0.2373} & 0.2945 & 0.1312 & 0.1199 & 0.1152 & 0.2402 & 0.2399 & 0.2382 \\
        & \textbf{FairIB} & 0.1892 & 0.2402 & 0.2681 & 0.1512  & 0.2341 & 0.2871 & 0.1321  & 0.1137 & \underline{0.0992} & 0.2319 & \underline{0.2325} & \underline{0.2317} \\
        & \textbf{FairCoRe} & \underline{0.1989} & \textbf{0.2474} & \underline{0.2769} & \underline{0.1574} & 0.2372 & 0.2946 & \underline{0.1219} & \underline{0.1129} & 0.1080 & \underline{0.2316} & 0.2346 & 0.2369 \\
        \cline{2-14}
        & \textbf{FairDDA} & \textbf{0.2001} & \underline{0.2470} & \textbf{0.2771} & 0.1569 & \textbf{0.2378} & \underline{0.2959} & \textbf{0.1023} & \textbf{0.0916} & \textbf{0.0874} & \textbf{0.2117} & \textbf{0.2120} & \textbf{0.2206} \\
        & $\mathbf{p}$\textbf{-value} & 2.3$\mathrm{e}{-2}$ & 3.4$\mathrm{e}{-1}$ & 2.8$\mathrm{e}{-2}$ & 6.8$\mathrm{e}{-1}$& 4.1$\mathrm{e}{-2}$ & 4.5$\mathrm{e}{-1}$& 1.7$\mathrm{e}{-4}$& 6.1$\mathrm{e}{-5}$ & 3.9$\mathrm{e}{-5}$& 3.1$\mathrm{e}{-3}$& 3.7$\mathrm{e}{-3}$ & 8.4$\mathrm{e}{-4}$ \\
        \hline
    \end{tabular}}
    \vspace{-.5em}
\end{table*}

\vspace{-.1em}
\subsubsection{\textbf{Evaluation metrics}}
\label{sec:metrics}

We use two widely used metrics, $Recall$ \cite{recall} and $NDCG$ \cite{ndcg} to evaluate the utility performance of recommendation tasks, and report the results with $\text{TopK}\in\{10,20,30\}$. In addition, to evaluate fairness, we adopt two metrics, $DP$ \cite{dwork2012dp} and $EO$ \cite{hardt2016eo}, which are respectively designed based on the group fairness definitions of \textit{Demographic Parity} and \textit{Equal Opportunity}. Formally, the metric $DP$ can be computed as follows:
\begin{equation}
\begin{aligned}
    &DP = JSD\left( P_{G_0}, P_{G_1} \right), \\
    &\text{where}
        \begin{cases}
        \; P_{G_s} = \left[ f_{G_s}(v) \right]_{v \in V}, \\[.8em]
        \; f_{G_s}(v) = \frac{\sum_{u \in G_s} \mathbb{I}(v \, \in \, \text{TopK}_u)}{|G_s|} 
    \end{cases}
    \hspace{-.5em}\text{for } s \in \{0,1\},
\end{aligned}
\end{equation}
where $JSD$ denotes the Jensen-Shannon Divergence \cite{lin2002divergence}, while $G_0$ and $G_1$ are the demographic groups with sensitive attributes $s = 0$ and $s = 1$, respectively. $\mathbb{I}(\cdot)$ is the indicator function, and $|G_s|$ is the number of users in group $G_s$. The metric $EO$ is defined as:
\begin{equation}
\begin{aligned}
    &EO = JSD\left( P_{G_0}, P_{G_1} \right), \\
    &\text{where}
    \begin{cases}
        \; P_{G_s} = \left[ d_{G_s}(v) \right]_{v \in V}, \\[.8em]
        \; d_{G_s}(v)=\frac{\sum_{u \in G_s} \mathbb{I}(v \, \in \, R_u \, \cap \, \text{TopK}_u)}{|G_s|} 
    \end{cases}
    \hspace{-.5em}\text{for } s \in \{0,1\},
\end{aligned}
\end{equation}
Note that higher $Recall$ and $NDCG$ indicate better utility, while lower $DP$ and $EO$ suggest better fairness.

\subsubsection{\textbf{Baselines}}

Several recent fair recommendation methods are compared with our method, including regularization-based, adversarial learning-based, data augmentation-based, disentanglement-based methods. The details of the baselines\footnote{It is worth noting that all baseline methods adopt \textbf{LightGCN} \cite{lightgcn} as the backbone, which is referred to as the base method in the subsequent experiments.} are given as follows:

\begin{itemize}[leftmargin=*]
\item \textbf{BP \cite{beyondparity}}: It introduces four unfairness metrics to address different forms of unfairness in recommendation and provides corresponding fairness objectives for fairness optimization.

\item \textbf{Adv \cite{adv}}: This work introduces an adversarial framework to enforce fairness constraints on graph embeddings, reducing relevance between sensitive attributes and the embeddings.

\item \textbf{FDA \cite{chen2023fda}}: It assumes that different user groups have similar preferences and generates fake interaction data based on this assumption to balance the training data, thus improving the recommendation fairness.

\item \textbf{FairRec \cite{fairrec}}: This work advocates that the data representation should be orthogonal to the sensitive component and constructs unbiased data representations through adversarial learning and orthogonal regularization.

\item \textbf{FairGo \cite{fairgo}}: It exploits the user-centred graph structure to learn the sensitive information filter based on the framework of adversarial learning to optimize the recommendation fairness.

\item \textbf{FairGNN \cite{fairgnn}}: It utilizes graph structure information and limited sensitive attribute information to improve the fairness through adversarial debiased learning and covariance constraints;

\item \textbf{FairMI \cite{fairmi}}: It introduces a dual MI optimization objective to capture non-sensitive information and disentangle sensitive information to construct fair data representations, which promotes fairness and maintains utility performance of recommendation. 

\item \textbf{FairIB \cite{fairib}}: This work frames the fairness task from the perspective of the Information Bottleneck principle, and employs a mutual information minimization method based on HSIC\cite{gretton2005HSIC} to learn fair user representations away from sensitive information, thereby improving recommendation fairness.

\item \textbf{FairCoRe \cite{faircore}}: It constructs two bias-aware representations based on the real-world scenarios of user attributes and the counterfactual scenarios where user-sensitive attributes are altered. It then employs mutual information optimization and an attribute-neutralization strategy to learn attribute-invariant fair representations, aiming to improve recommendation fairness.
\end{itemize}

\subsection{Overall performance (RQ1)}
In Table \ref{tab:performance}, we present the overall performance of our method compared with all baselines. The best result is emphasized in \textbf{bold}, and the second-best result is highlighted with an \underline{underline}. In addition, we provide the $p$-values to validate the significant performance improvements achieved by our model compared to the best baseline methods. Based on the results, we draw several observations:

\begin{itemize}[leftmargin=*]
\item Our method consistently achieves superior fairness, as measured by $DP$ and $EO$ metrics, with all improvements being statistically significant ($p<0.05$), effectively demonstrating the effectiveness of our method in improving recommendation fairness. Notably, our approach achieves reductions of approximately 3.4\% in $DP@10$ and 7.6\% in $EO@10$ on ML-1M, and 16.1\% in $DP@10$ and 8.6\% in $EO@10$ on LastFM, compared to the second-best methods. Most existing fair recommendation methods (e.g. FairIB and FairCoRe) focus on designing fairness objectives but neglect the effects of unbalanced data on representation learning, which results in limited fairness improvement. Our method design two data augmentation strategies to enhance data diversity and alleviate the bias arising from data imbalance and sensitive attributes, thus enhancing the fairness of recommendations.

\item Despite being primarily designed to improve fairness, FairDDA still achieves slight performance gains in terms of utility compared to LightGCN. This is because our method uses graph contrastive learning to preserve the original graph information, while eliminating bias related to sensitive attributes in the data by pruning sensitive edges and masking biased feature dimensions. This design disrupts the feature homogeneity inherent in traditional GCNs, effectively capturing user-personalized item preferences.

\item Although the existing data augmentation method (i.e., FDA) improves fairness to some extent, it incurs significant utility loss. This is because it enforces symmetric interaction behaviors across demographic groups, ignoring the inherent preference differences among users within those groups. In contrast, our method uses graph contrastive learning to ensure consistency between the original and augmented graph data, allowing the model to balance data distribution while preserving genuine user preferences. Additionally, our approach provides a clear definition of sensitive information based on two prior hypotheses, allowing the model to effectively identify and mitigate sensitivity, resulting in fairer representations. With these designs, FairDDA achieves an optimal balance between recommendation accuracy and fairness.
\end{itemize}

\subsection{Ablation Study (RQ2)}

\begin{table}[t]
    \caption{Results of ablation experiments at TopK=10.}
    \label{tab:ablation}
    \centering
    \resizebox{\linewidth}{!}{
    \begin{tabular}{c|c|c|c|c|c}
    \hline
    \multirow{1}{*}{\textbf{Dataset}} & \multirow{1}{*}{\textbf{Variant}} & \multicolumn{1}{c|}{\textit{\textbf{NDCG}}} & \multicolumn{1}{c|}{\textit{\textbf{Recall}}} & \multicolumn{1}{c|}{\textit{\textbf{DP}}} & \multicolumn{1}{c}{\textit{\textbf{EO}}} \\
    \hline
    \hline
    \multirow{5}{*}{\textbf{ML-1M}} & \textbf{LightGCN} & 0.2018  & 0.1511 & 0.2919 & 0.3609 \\
    \cline{2-6}
    & \textbf{FairDDA-}$\boldsymbol{\mathcal{L}_{dl}}$ & \textbf{0.2156} & \textbf{0.1741} & 0.2462 & 0.3018 \\
    & \textbf{FairDDA-EP} & \underline{0.2140}  & \underline{0.1606} & \underline{0.1251} & \underline{0.1909} \\
    & \textbf{FairDDA-FM} & 0.2073 & 0.1563 & 0.1291 & 0.1953 \\
    & \textbf{FairDDA} & 0.2138  & 0.1601 & \textbf{0.1183} & \textbf{0.1833} \\
    \hline
    \multirow{5}{*}{\textbf{LastFM}} & \textbf{LightGCN} & 0.1971 & 0.1572 & 0.2860 & 0.3508 \\
    \cline{2-6}
    & \textbf{FairDDA-}$\boldsymbol{\mathcal{L}_{dl}}$ & \textbf{0.2073} & \textbf{0.1688} & 0.2235 & 0.2941 \\
    & \textbf{FairDDA-EP} & 0.1980 & 0.1571 & \underline{0.1042} & \underline{0.2158} \\
    & \textbf{FairDDA-FM} & \underline{0.2004} & \underline{0.1574} & 0.1163 & 0.2231 \\
    & \textbf{FairDDA} & 0.2001 & 0.1569 & \textbf{0.1023} & \textbf{0.2117} \\
    \hline
    \end{tabular}}
    \vspace{-1em}
\end{table}

In this study, we propose two data augmentation strategies to mitigate sensitive information and learn fair feature representations. To evaluate their effectiveness, we conduct ablation studies by individually removing each strategy or its corresponding loss from the model, resulting in the following three variants:

\begin{itemize}[leftmargin=*]
\item \textbf{FairDDA-}$\boldsymbol{\mathcal{L}_{dl}}$ disables the debiasing learning.
\item \textbf{FairDDA-EP} disables the sensitive edge pruning strategy, 
\item \textbf{FairDDA-FM} disables the sensitive feature masking strategy.
\end{itemize}

As shown in Table~\ref{tab:ablation}, the results indicate that: i) \textbf{FairDDA} consistently achieves the best fairness performance across all datasets, demonstrating the effectiveness of the proposed strategies in enhancing recommendation fairness. ii) \textbf{FairDDA-}$\boldsymbol{\mathcal{L}_{dl}}$ achieves the largest utility gain over LightGCN but almost entirely loses fairness. This is because debiasing learning is designed to improve fairness by excluding sensitive information from the augmented graph. Once disabled, graph contrastive learning is no longer constrained by fairness objectives. As a result, the augmented graph acts merely as an information enrichment mechanism that alleviates data sparsity and improves utility, but fails to ensure fair representation learning. iii) \textbf{FairDDA-EP} consistently reduces fairness with only a minor impact on utility, indicating that the sensitive edge pruning strategy effectively mitigates bias while preserving user preference learning. iv) \textbf{FairDDA-FM} notably reduces fairness across all datasets, underscoring the importance of sensitive feature masking. However, it reduces utility on ML-1M but boosts it on LastFM. We attribute this to dataset traits: on the denser ML-1M, the model can more accurately distinguish sensitive features from non-sensitive ones, whereas the sparsity of LastFM limits the strategy’s effectiveness, as the learned representations lack sufficient supervised signals.

\vspace{-.3em}
\subsection{Hyperparameter Analysis (RQ3)}

\begin{figure}[!h]
    \centering
    \vspace{-1.5em}
    \subfloat[$\lambda_r$ \label{fig:lambda_r}]{%
        \includegraphics[width=0.33\linewidth]{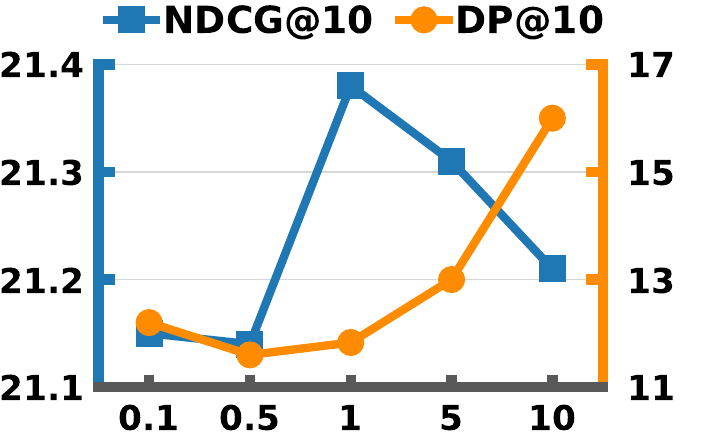}%
    } \hfill
    \subfloat[$\lambda_c$ \label{fig:lambda_c}]{%
        \includegraphics[width=0.33\linewidth]{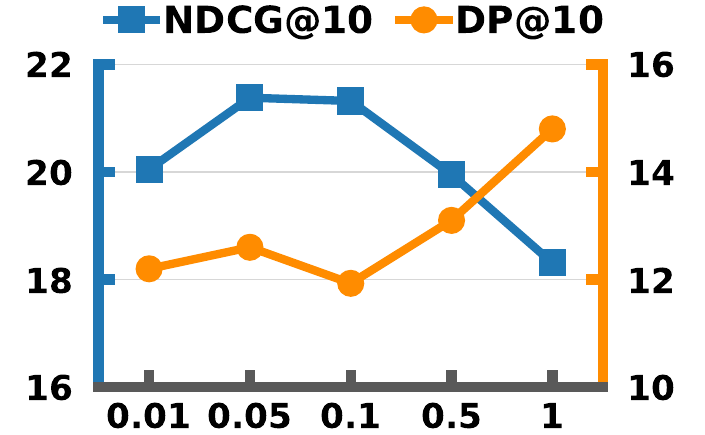}%
    } \hfill
    \subfloat[$\lambda_d$ \label{fig:lambda_d}]{%
        \includegraphics[width=0.33\linewidth]{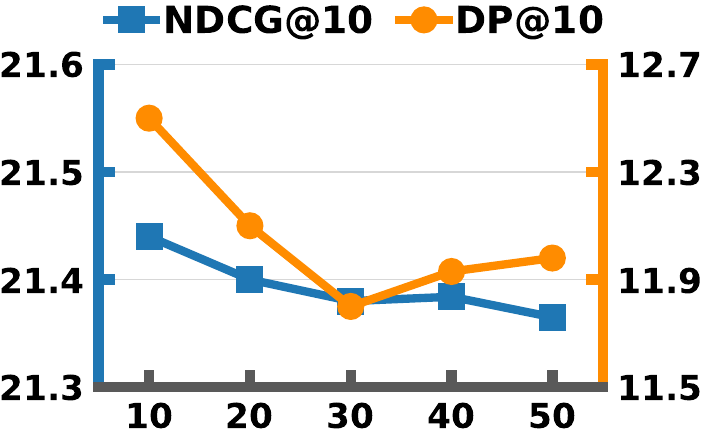}%
    }
    \vspace{-0.9em}
    \caption{Hyperparameter experiment on ML-1M, with all values on the y-axis scaled by a factor of 100.}
    \label{tab:Hyperparameter}
    \vspace{-0.5em}
\end{figure}

\noindent We conduct hyperparameter experiments to investigate the impact of the loss weights $\lambda_r$, $\lambda_c$, and $\lambda_d$ on both fairness and utility. As shown in Figure~\ref{tab:Hyperparameter}, the best trade-off between fairness and utility is achieved when $\lambda_r = 1$, $\lambda_c = 0.1$, and $\lambda_d = 30$. In addition, we observe that increasing $\lambda_r$ and $\lambda_c$ (e.g., $\lambda_r \in [1.0, 5.0]$, $\lambda_c \in [0.05, 0.1]$) encourages the model to better preserve the informativeness of augmented data, thereby maintaining utility performance. However, excessively high values of $\lambda_c$ (e.g., $\lambda_c \in [0.5, 1.0]$) would hinder the model’s ability to learn useful information from the training data. Conversely, lower values of $\lambda_r$ and $\lambda_c$ promote greater data diversity and fairness, but at the cost of information loss. For example, while the model achieves better fairness with $\lambda_r = 0.5$, its utility performance drops significantly compared to that with $\lambda_r = 1.0$. Furthermore, increasing $\lambda_d$ generally leads to decreased utility, whereas a moderate value (i.e., $\lambda_d = 30$) offers the best balance between fairness and utility.

\begin{figure}[t]
    \centering
    \subfloat[Random \label{fig:random}]{%
        \includegraphics[width=0.33\linewidth]{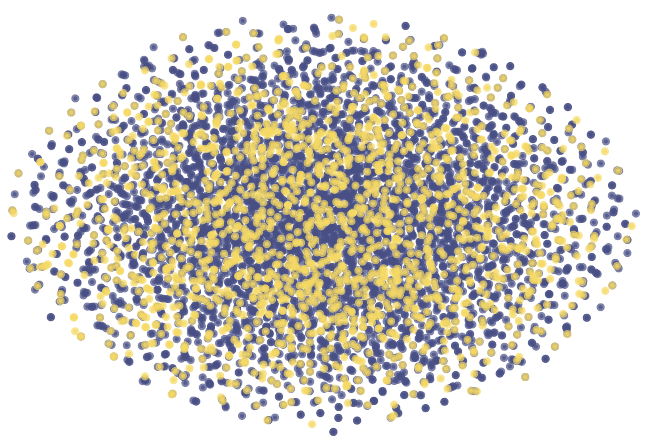}%
    } \hfill
    \subfloat[Base (LightGCN) \label{fig:base}]{%
        \includegraphics[width=0.33\linewidth]{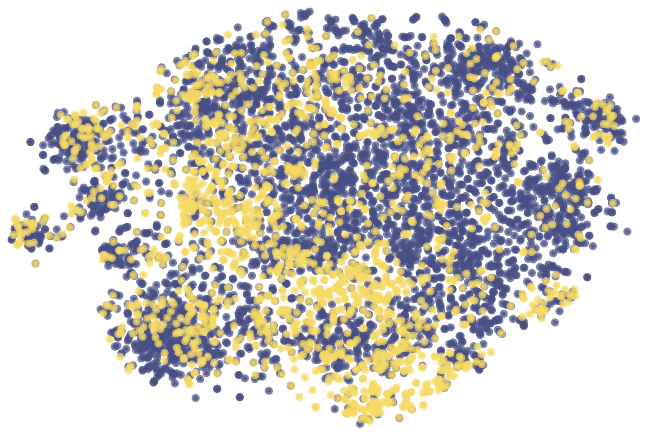}%
    } \hfill
    \subfloat[FDA \label{fig:fda}]{%
        \includegraphics[width=0.33\linewidth]{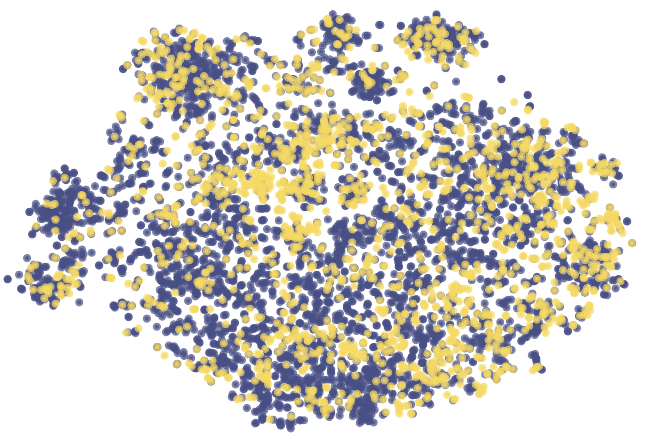}%
    } \hfill
    \subfloat[FairIB \label{fig:fairib}]{%
        \includegraphics[width=0.33\linewidth]{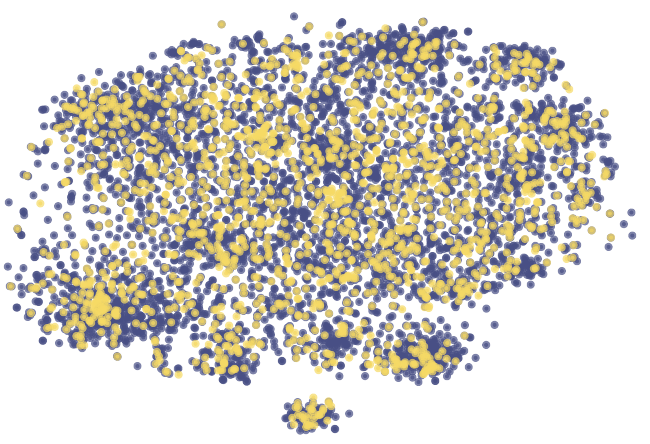}%
    } \hfill
    \subfloat[FairCoRe \label{fig:fairmi}]{%
        \includegraphics[width=0.33\linewidth]{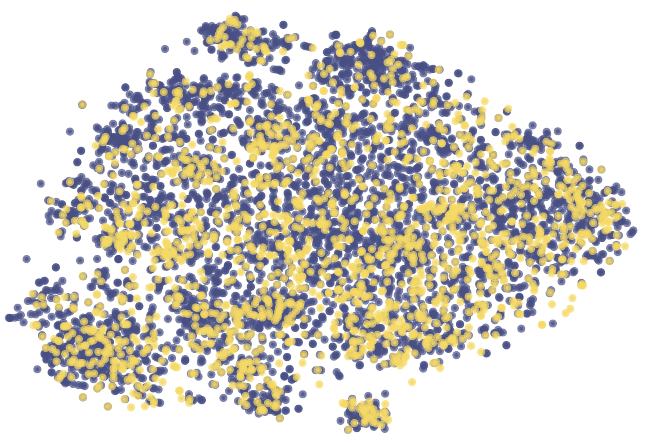}%
    } \hfill
    \subfloat[FairDDA \label{fig:fairdda}]{%
        \includegraphics[width=0.33\linewidth]{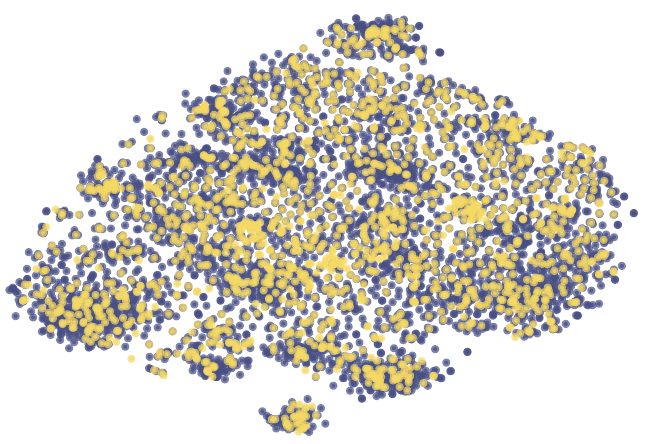}%
    }%
    \caption{Visualization of different methods w.r.t. the gender attribute on ML-1M, where user points are colored with two colors to indicate their respective gender groups.}
    \label{fig:visualization}
    \vspace{-1.5em}
\end{figure}

\subsection{Visualization (RQ4)}

To effectively show the fairness advantage of the proposed method, we employ t-SNE \cite{van2008tsne} to project user representations into a 2-dimensional space, where each user point is colored based on the gender attribute. Additionally, we compare the results with a randomly initialized model (denoted as \textbf{Random} in the figure), which is theoretically expected to exhibit the highest level of fairness. As shown in Figure~\ref{fig:visualization}, the visualization reveals the following insights:

\begin{itemize}[leftmargin=*]
\item \textbf{Random}: Due to randomly initialized representations, user embeddings are uniformly scattered in the feature space, showing no gender bias, but they fail to capture any user preference patterns.

\item \textbf{Base} and \textbf{FDA}: User representations learned by the model can be clearly separated by gender, indicating that the model strongly depends on gender features and introduces bias.

\item \textbf{FairIB}: User representations are nearly uniformly distributed, indicating strong fairness performance, but the lack of clear clusters suggests limited ability to capture user preferences.

\item \textbf{FairDDA} and \textbf{FairCoRe}: Both models form many clear representation clusters, indicating effective modeling for users with similar preferences and strong recommendation accuracy. Furthermore, in our model, users from different gender groups show more overlaps and are less distinguishable by gender compared to FairCoRe, highlighting the superior fairness of our method.
\end{itemize}

\section{Conclusion}

In this work, we present a dual data augmentation framework, which aims to establish balanced data distribution to train a fair and effective recommendation model. Specifically, based on the input user-item interaction graph and feature representations, we design two augmentation strategies, pruning sensitive edges and masking sensitive features, to construct a fair augmented graph and its representations. To achieve the optimal balance between fairness and utility, we encourage the consistency between original and augmented graphs, and minimize the dependence between augmented graph representations and sensitive information based on Hilbert-Schmidt Independence Criterion. Extensive experiments on two datasets show that our method improves both fairness and utility. In the future, we will explore additional data augmentation methods for fair recommendation task, such as inserting fairness-enhancing edges into the user-item interaction graph.

\begin{acks}
This work was supported in part by the National Natural Science Foundation of China (No. U22A2099), the Guangxi Science and Technology Base and Talent Special Project (No. AD23026268), and the Innovation Project of GUET Graduate Education (No. 2025YCXS058).
\end{acks}

\section*{GenAI Usage Disclosure}

The authors used generative AI tools (e.g., ChatGPT) solely for improving the grammar and clarity of the text. No content, analysis, code, or ideas generated by AI tools were used in the conception, development, or evaluation of the research.

\balance

\bibliographystyle{ACM-Reference-Format}
\bibliography{references}


\begin{thebibliography}{50}


\ifx \showCODEN    \undefined \def \showCODEN     #1{\unskip}     \fi
\ifx \showISBNx    \undefined \def \showISBNx     #1{\unskip}     \fi
\ifx \showISBNxiii \undefined \def \showISBNxiii  #1{\unskip}     \fi
\ifx \showISSN     \undefined \def \showISSN      #1{\unskip}     \fi
\ifx \showLCCN     \undefined \def \showLCCN      #1{\unskip}     \fi
\ifx \shownote     \undefined \def \shownote      #1{#1}          \fi
\ifx \showarticletitle \undefined \def \showarticletitle #1{#1}   \fi
\ifx \showURL      \undefined \def \showURL       {\relax}        \fi
\providecommand\bibfield[2]{#2}
\providecommand\bibinfo[2]{#2}
\providecommand\natexlab[1]{#1}
\providecommand\showeprint[2][]{arXiv:#2}

\bibitem[Agarwal et~al\mbox{.}(2021)]%
        {agarwal2021nifty}
\bibfield{author}{\bibinfo{person}{Chirag Agarwal}, \bibinfo{person}{Himabindu Lakkaraju}, {and} \bibinfo{person}{Marinka Zitnik}.} \bibinfo{year}{2021}\natexlab{}.
\newblock \showarticletitle{Towards a unified framework for fair and stable graph representation learning}. In \bibinfo{booktitle}{\emph{UAI}}. PMLR, \bibinfo{pages}{2114--2124}.
\newblock


\bibitem[Bechavod et~al\mbox{.}(2020)]%
        {individualfairness}
\bibfield{author}{\bibinfo{person}{Yahav Bechavod}, \bibinfo{person}{Christopher Jung}, {and} \bibinfo{person}{Steven~Z Wu}.} \bibinfo{year}{2020}\natexlab{}.
\newblock \showarticletitle{Metric-free individual fairness in online learning}. In \bibinfo{booktitle}{\emph{NeurIPS}}. \bibinfo{pages}{11214--11225}.
\newblock


\bibitem[Bin et~al\mbox{.}(2025)]%
        {faircore}
\bibfield{author}{\bibinfo{person}{Chenzhong Bin}, \bibinfo{person}{Wenqiang Liu}, \bibinfo{person}{Feng Zhang}, \bibinfo{person}{Liang Chang}, {and} \bibinfo{person}{Tianlong Gu}.} \bibinfo{year}{2025}\natexlab{}.
\newblock \showarticletitle{FairCoRe: Fairness-aware Recommendation through Counterfactual Representation Learning}.
\newblock \bibinfo{journal}{\emph{IEEE Trans. Knowl. Data Eng.}} (\bibinfo{year}{2025}).
\newblock


\bibitem[Bose and Hamilton(2019)]%
        {adv}
\bibfield{author}{\bibinfo{person}{Avishek Bose} {and} \bibinfo{person}{William Hamilton}.} \bibinfo{year}{2019}\natexlab{}.
\newblock \showarticletitle{Compositional fairness constraints for graph embeddings}. In \bibinfo{booktitle}{\emph{ICML}}. PMLR, \bibinfo{pages}{715--724}.
\newblock


\bibitem[Burke et~al\mbox{.}(2018)]%
        {burke2018balanced}
\bibfield{author}{\bibinfo{person}{Robin Burke}, \bibinfo{person}{Nasim Sonboli}, {and} \bibinfo{person}{Aldo Ordonez-Gauger}.} \bibinfo{year}{2018}\natexlab{}.
\newblock \showarticletitle{Balanced neighborhoods for multi-sided fairness in recommendation}. In \bibinfo{booktitle}{\emph{FAT}}. PMLR, \bibinfo{pages}{202--214}.
\newblock


\bibitem[Casella and Berger(2024)]%
        {statistical}
\bibfield{author}{\bibinfo{person}{George Casella} {and} \bibinfo{person}{Roger Berger}.} \bibinfo{year}{2024}\natexlab{}.
\newblock \bibinfo{booktitle}{\emph{Statistical inference}}.
\newblock \bibinfo{publisher}{Chapman and Hall/CRC}.
\newblock


\bibitem[Celma~Herrada et~al\mbox{.}(2009)]%
        {LastFM}
\bibfield{author}{\bibinfo{person}{{\`O}scar Celma~Herrada} {et~al\mbox{.}}} \bibinfo{year}{2009}\natexlab{}.
\newblock \bibinfo{booktitle}{\emph{Music recommendation and discovery in the long tail}}.
\newblock \bibinfo{publisher}{Universitat Pompeu Fabra}.
\newblock


\bibitem[Chen et~al\mbox{.}(2023)]%
        {chen2023fda}
\bibfield{author}{\bibinfo{person}{Lei Chen}, \bibinfo{person}{Le Wu}, \bibinfo{person}{Kun Zhang}, \bibinfo{person}{Richang Hong}, \bibinfo{person}{Defu Lian}, \bibinfo{person}{Zhiqiang Zhang}, \bibinfo{person}{Jun Zhou}, {and} \bibinfo{person}{Meng Wang}.} \bibinfo{year}{2023}\natexlab{}.
\newblock \showarticletitle{Improving recommendation fairness via data augmentation}. In \bibinfo{booktitle}{\emph{WWW}}. \bibinfo{pages}{1012--1020}.
\newblock


\bibitem[Chen et~al\mbox{.}(2024)]%
        {chen2024fairgap}
\bibfield{author}{\bibinfo{person}{Wei Chen}, \bibinfo{person}{Yiqing Wu}, \bibinfo{person}{Zhao Zhang}, \bibinfo{person}{Fuzhen Zhuang}, \bibinfo{person}{Zhongshi He}, \bibinfo{person}{Ruobing Xie}, {and} \bibinfo{person}{Feng Xia}.} \bibinfo{year}{2024}\natexlab{}.
\newblock \showarticletitle{FairGap: Fairness-aware recommendation via generating counterfactual graph}.
\newblock \bibinfo{journal}{\emph{ACM Trans. Inf. Syst.}} \bibinfo{volume}{42}, \bibinfo{number}{4} (\bibinfo{year}{2024}), \bibinfo{pages}{1--25}.
\newblock


\bibitem[Cheng et~al\mbox{.}(2021)]%
        {fairfil}
\bibfield{author}{\bibinfo{person}{Pengyu Cheng}, \bibinfo{person}{Weituo Hao}, \bibinfo{person}{Siyang Yuan}, \bibinfo{person}{Shijing Si}, {and} \bibinfo{person}{Lawrence Carin}.} \bibinfo{year}{2021}\natexlab{}.
\newblock \showarticletitle{FairFil: Contrastive Neural Debiasing Method for Pretrained Text Encoders}. In \bibinfo{booktitle}{\emph{ICLR}}.
\newblock


\bibitem[Dai and Wang(2021)]%
        {fairgnn}
\bibfield{author}{\bibinfo{person}{Enyan Dai} {and} \bibinfo{person}{Suhang Wang}.} \bibinfo{year}{2021}\natexlab{}.
\newblock \showarticletitle{Say no to the discrimination: Learning fair graph neural networks with limited sensitive attribute information}. In \bibinfo{booktitle}{\emph{WSDM}}. \bibinfo{pages}{680--688}.
\newblock


\bibitem[Deldjoo et~al\mbox{.}(2024)]%
        {summarize1}
\bibfield{author}{\bibinfo{person}{Yashar Deldjoo}, \bibinfo{person}{Dietmar Jannach}, \bibinfo{person}{Alejandro Bellogin}, \bibinfo{person}{Alessandro Difonzo}, {and} \bibinfo{person}{Dario Zanzonelli}.} \bibinfo{year}{2024}\natexlab{}.
\newblock \showarticletitle{Fairness in recommender systems: research landscape and future directions}.
\newblock \bibinfo{journal}{\emph{User Model. User-Adapt. Interact.}} \bibinfo{volume}{34}, \bibinfo{number}{1} (\bibinfo{year}{2024}), \bibinfo{pages}{59--108}.
\newblock


\bibitem[D'Inc{\`a} et~al\mbox{.}(2024)]%
        {image_da}
\bibfield{author}{\bibinfo{person}{Moreno D'Inc{\`a}}, \bibinfo{person}{Christos Tzelepis}, \bibinfo{person}{Ioannis Patras}, {and} \bibinfo{person}{Nicu Sebe}.} \bibinfo{year}{2024}\natexlab{}.
\newblock \showarticletitle{Improving fairness using vision-language driven image augmentation}. In \bibinfo{booktitle}{\emph{WACV}}. \bibinfo{pages}{4695--4704}.
\newblock


\bibitem[Du et~al\mbox{.}(2020)]%
        {du2020fairness}
\bibfield{author}{\bibinfo{person}{Mengnan Du}, \bibinfo{person}{Fan Yang}, \bibinfo{person}{Na Zou}, {and} \bibinfo{person}{Xia Hu}.} \bibinfo{year}{2020}\natexlab{}.
\newblock \showarticletitle{Fairness in deep learning: A computational perspective}.
\newblock \bibinfo{journal}{\emph{IEEE Intell. Syst.}} \bibinfo{volume}{36}, \bibinfo{number}{4} (\bibinfo{year}{2020}), \bibinfo{pages}{25--34}.
\newblock


\bibitem[Dwork et~al\mbox{.}(2012)]%
        {dwork2012dp}
\bibfield{author}{\bibinfo{person}{Cynthia Dwork}, \bibinfo{person}{Moritz Hardt}, \bibinfo{person}{Toniann Pitassi}, \bibinfo{person}{Omer Reingold}, {and} \bibinfo{person}{Richard Zemel}.} \bibinfo{year}{2012}\natexlab{}.
\newblock \showarticletitle{Fairness through awareness}. In \bibinfo{booktitle}{\emph{ITCS}}. \bibinfo{pages}{214--226}.
\newblock


\bibitem[Ekstrand et~al\mbox{.}(2018)]%
        {pmlr-v81-ekstrand18b}
\bibfield{author}{\bibinfo{person}{Michael~D Ekstrand}, \bibinfo{person}{Mucun Tian}, \bibinfo{person}{Ion~Madrazo Azpiazu}, \bibinfo{person}{Jennifer~D Ekstrand}, \bibinfo{person}{Oghenemaro Anuyah}, \bibinfo{person}{David McNeill}, {and} \bibinfo{person}{Maria~Soledad Pera}.} \bibinfo{year}{2018}\natexlab{}.
\newblock \showarticletitle{All the cool kids, how do they fit in?: Popularity and demographic biases in recommender evaluation and effectiveness}. In \bibinfo{booktitle}{\emph{FAT}}. PMLR, \bibinfo{pages}{172--186}.
\newblock


\bibitem[Gretton et~al\mbox{.}(2005)]%
        {gretton2005HSIC}
\bibfield{author}{\bibinfo{person}{Arthur Gretton}, \bibinfo{person}{Olivier Bousquet}, \bibinfo{person}{Alex Smola}, {and} \bibinfo{person}{Bernhard Sch{\"o}lkopf}.} \bibinfo{year}{2005}\natexlab{}.
\newblock \showarticletitle{Measuring statistical dependence with Hilbert-Schmidt norms}. In \bibinfo{booktitle}{\emph{ALT}}. Springer, \bibinfo{pages}{63--77}.
\newblock


\bibitem[Gunawardana and Shani(2009)]%
        {recall}
\bibfield{author}{\bibinfo{person}{Asela Gunawardana} {and} \bibinfo{person}{Guy Shani}.} \bibinfo{year}{2009}\natexlab{}.
\newblock \showarticletitle{A Survey of Accuracy Evaluation Metrics of Recommendation Tasks}.
\newblock \bibinfo{journal}{\emph{J. Mach. Learn. Res.}} \bibinfo{volume}{10}, \bibinfo{number}{12} (\bibinfo{year}{2009}).
\newblock


\bibitem[Guo et~al\mbox{.}(2022)]%
        {guo2022distance}
\bibfield{author}{\bibinfo{person}{Dandan Guo}, \bibinfo{person}{Chaojie Wang}, \bibinfo{person}{Baoxiang Wang}, {and} \bibinfo{person}{Hongyuan Zha}.} \bibinfo{year}{2022}\natexlab{}.
\newblock \showarticletitle{Learning fair representations via distance correlation minimization}.
\newblock \bibinfo{journal}{\emph{IEEE Trans. Neural Netw.}} \bibinfo{volume}{35}, \bibinfo{number}{2} (\bibinfo{year}{2022}), \bibinfo{pages}{2139--2152}.
\newblock


\bibitem[Ham et~al\mbox{.}(2004)]%
        {ham2004kernel}
\bibfield{author}{\bibinfo{person}{Jihun Ham}, \bibinfo{person}{Daniel~D Lee}, \bibinfo{person}{Sebastian Mika}, {and} \bibinfo{person}{Bernhard Sch{\"o}lkopf}.} \bibinfo{year}{2004}\natexlab{}.
\newblock \showarticletitle{A kernel view of the dimensionality reduction of manifolds}. In \bibinfo{booktitle}{\emph{ICML}}. \bibinfo{pages}{47}.
\newblock


\bibitem[Hardt et~al\mbox{.}(2016)]%
        {hardt2016eo}
\bibfield{author}{\bibinfo{person}{Moritz Hardt}, \bibinfo{person}{Eric Price}, {and} \bibinfo{person}{Nati Srebro}.} \bibinfo{year}{2016}\natexlab{}.
\newblock \showarticletitle{Equality of opportunity in supervised learning}. In \bibinfo{booktitle}{\emph{NeurIPS}}. \bibinfo{pages}{3315--3323}.
\newblock


\bibitem[Harper and Konstan(2015)]%
        {movielens}
\bibfield{author}{\bibinfo{person}{F~Maxwell Harper} {and} \bibinfo{person}{Joseph~A Konstan}.} \bibinfo{year}{2015}\natexlab{}.
\newblock \showarticletitle{The movielens datasets: History and context}.
\newblock \bibinfo{journal}{\emph{ACM Trans. Interact. Intell. Syst.}} \bibinfo{volume}{5}, \bibinfo{number}{4} (\bibinfo{year}{2015}), \bibinfo{pages}{19}.
\newblock


\bibitem[He et~al\mbox{.}(2020)]%
        {lightgcn}
\bibfield{author}{\bibinfo{person}{Xiangnan He}, \bibinfo{person}{Kuan Deng}, \bibinfo{person}{Xiang Wang}, \bibinfo{person}{Yan Li}, \bibinfo{person}{Yongdong Zhang}, {and} \bibinfo{person}{Meng Wang}.} \bibinfo{year}{2020}\natexlab{}.
\newblock \showarticletitle{Lightgcn: Simplifying and powering graph convolution network for recommendation}. In \bibinfo{booktitle}{\emph{SIGIR}}. \bibinfo{pages}{639--648}.
\newblock


\bibitem[Iosifidis et~al\mbox{.}(2019)]%
        {iosifidis2019fae}
\bibfield{author}{\bibinfo{person}{Vasileios Iosifidis}, \bibinfo{person}{Besnik Fetahu}, {and} \bibinfo{person}{Eirini Ntoutsi}.} \bibinfo{year}{2019}\natexlab{}.
\newblock \showarticletitle{Fae: A fairness-aware ensemble framework}. In \bibinfo{booktitle}{\emph{IEEE BigData}}. IEEE, \bibinfo{pages}{1375--1380}.
\newblock


\bibitem[Jang et~al\mbox{.}(2017)]%
        {Jang2017gumbel}
\bibfield{author}{\bibinfo{person}{Eric Jang}, \bibinfo{person}{Shixiang Gu}, {and} \bibinfo{person}{Ben Poole}.} \bibinfo{year}{2017}\natexlab{}.
\newblock \showarticletitle{Categorical Reparametrization with Gumble-Softmax}. In \bibinfo{booktitle}{\emph{ICLR}}.
\newblock


\bibitem[J{\"a}rvelin and Kek{\"a}l{\"a}inen(2000)]%
        {ndcg}
\bibfield{author}{\bibinfo{person}{Kalervo J{\"a}rvelin} {and} \bibinfo{person}{Jaana Kek{\"a}l{\"a}inen}.} \bibinfo{year}{2000}\natexlab{}.
\newblock \showarticletitle{IR evaluation methods for retrieving highly relevant documents.}. In \bibinfo{booktitle}{\emph{SIGIR}}. ACM, \bibinfo{pages}{41--48}.
\newblock


\bibitem[Kose and Shen(2022)]%
        {kose2022fairaug}
\bibfield{author}{\bibinfo{person}{O~Deniz Kose} {and} \bibinfo{person}{Yanning Shen}.} \bibinfo{year}{2022}\natexlab{}.
\newblock \showarticletitle{Fair node representation learning via adaptive data augmentation}.
\newblock \bibinfo{journal}{\emph{arXiv:2201.08549}} (\bibinfo{year}{2022}).
\newblock


\bibitem[Li et~al\mbox{.}(2021)]%
        {li2021user}
\bibfield{author}{\bibinfo{person}{Yunqi Li}, \bibinfo{person}{Hanxiong Chen}, \bibinfo{person}{Zuohui Fu}, \bibinfo{person}{Yingqiang Ge}, {and} \bibinfo{person}{Yongfeng Zhang}.} \bibinfo{year}{2021}\natexlab{}.
\newblock \showarticletitle{User-oriented fairness in recommendation}. In \bibinfo{booktitle}{\emph{WWW}}. \bibinfo{pages}{624--632}.
\newblock


\bibitem[Li et~al\mbox{.}(2022)]%
        {summarize2}
\bibfield{author}{\bibinfo{person}{Yunqi Li}, \bibinfo{person}{Hanxiong Chen}, \bibinfo{person}{Shuyuan Xu}, \bibinfo{person}{Yingqiang Ge}, \bibinfo{person}{Juntao Tan}, \bibinfo{person}{Shuchang Liu}, {and} \bibinfo{person}{Yongfeng Zhang}.} \bibinfo{year}{2022}\natexlab{}.
\newblock \showarticletitle{Fairness in recommendation: A survey}.
\newblock \bibinfo{journal}{\emph{arXiv:2205.13619}} (\bibinfo{year}{2022}).
\newblock


\bibitem[Lin(2002)]%
        {lin2002divergence}
\bibfield{author}{\bibinfo{person}{Jianhua Lin}.} \bibinfo{year}{2002}\natexlab{}.
\newblock \showarticletitle{Divergence measures based on the Shannon entropy}.
\newblock \bibinfo{journal}{\emph{IEEE Trans. Inf. Theory}} \bibinfo{volume}{37}, \bibinfo{number}{1} (\bibinfo{year}{2002}), \bibinfo{pages}{145--151}.
\newblock


\bibitem[Ling et~al\mbox{.}(2023)]%
        {ling2023learning}
\bibfield{author}{\bibinfo{person}{Hongyi Ling}, \bibinfo{person}{Zhimeng Jiang}, \bibinfo{person}{Youzhi Luo}, \bibinfo{person}{Shuiwang Ji}, {and} \bibinfo{person}{Na Zou}.} \bibinfo{year}{2023}\natexlab{}.
\newblock \showarticletitle{Learning fair graph representations via automated data augmentations}. In \bibinfo{booktitle}{\emph{ICLR}}.
\newblock


\bibitem[Ma et~al\mbox{.}(2022)]%
        {ma2022graphcounterfactual}
\bibfield{author}{\bibinfo{person}{Jing Ma}, \bibinfo{person}{Ruocheng Guo}, \bibinfo{person}{Mengting Wan}, \bibinfo{person}{Longqi Yang}, \bibinfo{person}{Aidong Zhang}, {and} \bibinfo{person}{Jundong Li}.} \bibinfo{year}{2022}\natexlab{}.
\newblock \showarticletitle{Learning fair node representations with graph counterfactual fairness}. In \bibinfo{booktitle}{\emph{WSDM}}. \bibinfo{pages}{695--703}.
\newblock


\bibitem[Ma et~al\mbox{.}(2020)]%
        {ma2020hsic}
\bibfield{author}{\bibinfo{person}{Wan-Duo~Kurt Ma}, \bibinfo{person}{JP Lewis}, {and} \bibinfo{person}{W~Bastiaan Kleijn}.} \bibinfo{year}{2020}\natexlab{}.
\newblock \showarticletitle{The HSIC bottleneck: Deep learning without back-propagation}. In \bibinfo{booktitle}{\emph{AAAI}}, Vol.~\bibinfo{volume}{34}. \bibinfo{pages}{5085--5092}.
\newblock


\bibitem[Oh et~al\mbox{.}(2022)]%
        {oh2022distributional}
\bibfield{author}{\bibinfo{person}{Changdae Oh}, \bibinfo{person}{Heeji Won}, \bibinfo{person}{Junhyuk So}, \bibinfo{person}{Taero Kim}, \bibinfo{person}{Yewon Kim}, \bibinfo{person}{Hosik Choi}, {and} \bibinfo{person}{Kyungwoo Song}.} \bibinfo{year}{2022}\natexlab{}.
\newblock \showarticletitle{Learning fair representation via distributional contrastive disentanglement}. In \bibinfo{booktitle}{\emph{SIGKDD}}. \bibinfo{pages}{1295--1305}.
\newblock


\bibitem[Pedreschi et~al\mbox{.}(2009)]%
        {groupfairnesssiam}
\bibfield{author}{\bibinfo{person}{Dino Pedreschi}, \bibinfo{person}{Salvatore Ruggieri}, {and} \bibinfo{person}{Franco Turini}.} \bibinfo{year}{2009}\natexlab{}.
\newblock \showarticletitle{Measuring discrimination in socially-sensitive decision records}. In \bibinfo{booktitle}{\emph{SDM}}. SIAM, \bibinfo{pages}{581--592}.
\newblock


\bibitem[Qi et~al\mbox{.}(2021)]%
        {social-media}
\bibfield{author}{\bibinfo{person}{Tao Qi}, \bibinfo{person}{Fangzhao Wu}, \bibinfo{person}{Chuhan Wu}, {and} \bibinfo{person}{Yongfeng Huang}.} \bibinfo{year}{2021}\natexlab{}.
\newblock \showarticletitle{Personalized news recommendation with knowledge-aware interactive matching}. In \bibinfo{booktitle}{\emph{SIGIR}}. \bibinfo{pages}{61--70}.
\newblock


\bibitem[Rendle et~al\mbox{.}(2009)]%
        {rendle2012bpr}
\bibfield{author}{\bibinfo{person}{Steffen Rendle}, \bibinfo{person}{Christoph Freudenthaler}, \bibinfo{person}{Zeno Gantner}, {and} \bibinfo{person}{Lars Schmidt-Thieme}.} \bibinfo{year}{2009}\natexlab{}.
\newblock \showarticletitle{BPR: Bayesian personalized ranking from implicit feedback}. In \bibinfo{booktitle}{\emph{UAI}}. \bibinfo{pages}{452--461}.
\newblock


\bibitem[Sarhan et~al\mbox{.}(2020)]%
        {sarhan2020orthogonal}
\bibfield{author}{\bibinfo{person}{Mhd~Hasan Sarhan}, \bibinfo{person}{Nassir Navab}, \bibinfo{person}{Abouzar Eslami}, {and} \bibinfo{person}{Shadi Albarqouni}.} \bibinfo{year}{2020}\natexlab{}.
\newblock \showarticletitle{Fairness by learning orthogonal disentangled representations}. In \bibinfo{booktitle}{\emph{ECCV}}. Springer, \bibinfo{pages}{746--761}.
\newblock


\bibitem[Shorten and Khoshgoftaar(2019)]%
        {shorten2019da_image}
\bibfield{author}{\bibinfo{person}{Connor Shorten} {and} \bibinfo{person}{Taghi~M Khoshgoftaar}.} \bibinfo{year}{2019}\natexlab{}.
\newblock \showarticletitle{A survey on image data augmentation for deep learning}.
\newblock \bibinfo{journal}{\emph{J. Big Data}} \bibinfo{volume}{6}, \bibinfo{number}{1} (\bibinfo{year}{2019}), \bibinfo{pages}{1--48}.
\newblock


\bibitem[Shorten et~al\mbox{.}(2021)]%
        {shorten2021da_text}
\bibfield{author}{\bibinfo{person}{Connor Shorten}, \bibinfo{person}{Taghi~M Khoshgoftaar}, {and} \bibinfo{person}{Borko Furht}.} \bibinfo{year}{2021}\natexlab{}.
\newblock \showarticletitle{Text data augmentation for deep learning}.
\newblock \bibinfo{journal}{\emph{J. Big Data}} \bibinfo{volume}{8}, \bibinfo{number}{1} (\bibinfo{year}{2021}), \bibinfo{pages}{101}.
\newblock


\bibitem[Van~der Maaten and Hinton(2008)]%
        {van2008tsne}
\bibfield{author}{\bibinfo{person}{Laurens Van~der Maaten} {and} \bibinfo{person}{Geoffrey Hinton}.} \bibinfo{year}{2008}\natexlab{}.
\newblock \showarticletitle{Visualizing data using t-SNE.}
\newblock \bibinfo{journal}{\emph{J. Mach. Learn. Res.}} \bibinfo{volume}{9}, \bibinfo{number}{11} (\bibinfo{year}{2008}).
\newblock


\bibitem[Wang et~al\mbox{.}(2018)]%
        {e-commerce}
\bibfield{author}{\bibinfo{person}{Jizhe Wang}, \bibinfo{person}{Pipei Huang}, \bibinfo{person}{Huan Zhao}, \bibinfo{person}{Zhibo Zhang}, \bibinfo{person}{Binqiang Zhao}, {and} \bibinfo{person}{Dik~Lun Lee}.} \bibinfo{year}{2018}\natexlab{}.
\newblock \showarticletitle{Billion-scale commodity embedding for e-commerce recommendation in alibaba}. In \bibinfo{booktitle}{\emph{SIGKDD}}. \bibinfo{pages}{839--848}.
\newblock


\bibitem[Wang et~al\mbox{.}(2023)]%
        {wang2023survey}
\bibfield{author}{\bibinfo{person}{Yifan Wang}, \bibinfo{person}{Weizhi Ma}, \bibinfo{person}{Min Zhang}, \bibinfo{person}{Yiqun Liu}, {and} \bibinfo{person}{Shaoping Ma}.} \bibinfo{year}{2023}\natexlab{}.
\newblock \showarticletitle{A survey on the fairness of recommender systems}.
\newblock \bibinfo{journal}{\emph{ACM Trans. Inf. Syst.}} \bibinfo{volume}{41}, \bibinfo{number}{3} (\bibinfo{year}{2023}), \bibinfo{pages}{1--43}.
\newblock


\bibitem[Wang et~al\mbox{.}(2022)]%
        {wang2022fairvgnn}
\bibfield{author}{\bibinfo{person}{Yu Wang}, \bibinfo{person}{Yuying Zhao}, \bibinfo{person}{Yushun Dong}, \bibinfo{person}{Huiyuan Chen}, \bibinfo{person}{Jundong Li}, {and} \bibinfo{person}{Tyler Derr}.} \bibinfo{year}{2022}\natexlab{}.
\newblock \showarticletitle{Improving fairness in graph neural networks via mitigating sensitive attribute leakage}. In \bibinfo{booktitle}{\emph{SIGKDD}}. \bibinfo{pages}{1938--1948}.
\newblock


\bibitem[Wu et~al\mbox{.}(2021b)]%
        {fairrec}
\bibfield{author}{\bibinfo{person}{Chuhan Wu}, \bibinfo{person}{Fangzhao Wu}, \bibinfo{person}{Xiting Wang}, \bibinfo{person}{Yongfeng Huang}, {and} \bibinfo{person}{Xing Xie}.} \bibinfo{year}{2021}\natexlab{b}.
\newblock \showarticletitle{Fairness-aware news recommendation with decomposed adversarial learning}. In \bibinfo{booktitle}{\emph{AAAI}}, Vol.~\bibinfo{volume}{35}. \bibinfo{pages}{4462--4469}.
\newblock


\bibitem[Wu et~al\mbox{.}(2021a)]%
        {fairgo}
\bibfield{author}{\bibinfo{person}{Le Wu}, \bibinfo{person}{Lei Chen}, \bibinfo{person}{Pengyang Shao}, \bibinfo{person}{Richang Hong}, \bibinfo{person}{Xiting Wang}, {and} \bibinfo{person}{Meng Wang}.} \bibinfo{year}{2021}\natexlab{a}.
\newblock \showarticletitle{Learning fair representations for recommendation: A graph-based perspective}. In \bibinfo{booktitle}{\emph{WWW}}. \bibinfo{pages}{2198--2208}.
\newblock


\bibitem[Xie et~al\mbox{.}(2024)]%
        {fairib}
\bibfield{author}{\bibinfo{person}{Junsong Xie}, \bibinfo{person}{Yonghui Yang}, \bibinfo{person}{Zihan Wang}, {and} \bibinfo{person}{Le Wu}.} \bibinfo{year}{2024}\natexlab{}.
\newblock \showarticletitle{Learning fair representations for recommendation via information bottleneck principle}. In \bibinfo{booktitle}{\emph{IJCAI}}. \bibinfo{pages}{2469--2477}.
\newblock


\bibitem[Yao and Huang(2017)]%
        {beyondparity}
\bibfield{author}{\bibinfo{person}{Sirui Yao} {and} \bibinfo{person}{Bert Huang}.} \bibinfo{year}{2017}\natexlab{}.
\newblock \showarticletitle{Beyond parity: Fairness objectives for collaborative filtering}. In \bibinfo{booktitle}{\emph{NeurIPS}}. \bibinfo{pages}{2921--2930}.
\newblock


\bibitem[Yu et~al\mbox{.}(2022)]%
        {simgcl}
\bibfield{author}{\bibinfo{person}{Junliang Yu}, \bibinfo{person}{Hongzhi Yin}, \bibinfo{person}{Xin Xia}, \bibinfo{person}{Tong Chen}, \bibinfo{person}{Lizhen Cui}, {and} \bibinfo{person}{Quoc Viet~Hung Nguyen}.} \bibinfo{year}{2022}\natexlab{}.
\newblock \showarticletitle{Are graph augmentations necessary? simple graph contrastive learning for recommendation}. In \bibinfo{booktitle}{\emph{SIGIR}}. \bibinfo{pages}{1294--1303}.
\newblock


\bibitem[Zhao et~al\mbox{.}(2023)]%
        {fairmi}
\bibfield{author}{\bibinfo{person}{Chen Zhao}, \bibinfo{person}{Le Wu}, \bibinfo{person}{Pengyang Shao}, \bibinfo{person}{Kun Zhang}, \bibinfo{person}{Richang Hong}, {and} \bibinfo{person}{Meng Wang}.} \bibinfo{year}{2023}\natexlab{}.
\newblock \showarticletitle{Fair representation learning for recommendation: a mutual information perspective}. In \bibinfo{booktitle}{\emph{AAAI}}, Vol.~\bibinfo{volume}{37}. \bibinfo{pages}{4911--4919}.
\newblock


\end{thebibliography}

\newpage
\appendix
\section{Evaluation in Multi-Class Sensitive Scenarios}
\label{sec:multi-sensitive}

\begin{table}[t]
    \caption{Performance comparison of all SOTA methods on the ML-1M dataset (occupation scenario) with TopK=10. ``N/A'' indicates methods inapplicable to the multi-class sensitive attribute setting.}
    \label{tab:4}
    \centering
    \resizebox{\linewidth}{!}{
    \begin{tabular}{c|c|c|c|c|c}
    \hline
    \multirow{1}{*}{\textbf{Dataset}} & \multirow{1}{*}{\textbf{Method}} & \multicolumn{1}{c|}{\textit{\textbf{NDCG}} $\uparrow$} & \multicolumn{1}{c|}{\textit{\textbf{Recall}} $\uparrow$} & \multicolumn{1}{c|}{\textit{\textbf{DP}} $\downarrow$} & \multicolumn{1}{c}{\textit{\textbf{EO}} $\downarrow$} \\
    \hline
    \hline
    \multirow{11}{*}{\textbf{ML-1M}} & \textbf{LightGCN} & 0.2018  & 0.1511 & 0.2919 & 0.3609 \\
    \cline{2-6}
    & \textbf{BP} & N/A & N/A & N/A & N/A \\
    & \textbf{ADV} & 0.1966 & 0.1457 & 0.2650 & 0.3693 \\
    & \textbf{FairRec} & 0.1957 & 0.1454 & 0.2707 & 0.3746 \\
    & \textbf{FDA} & 0.1913 & 0.1430 & 0.2951 & 0.3677 \\
    & \textbf{FairGo} & 0.1909 & 0.1426 & 0.2835 & 0.3559 \\
    & \textbf{FairGNN} & 0.1977 & 0.1480 & 0.2633 & 0.3629 \\
    & \textbf{FairIB} & 0.2105 & 0.1575 & \underline{0.2361} & \underline{0.3335} \\
    & \textbf{FairMI} & 0.2133 & 0.1591 & 0.2601 & 0.3616 \\
    & \textbf{FairCoRe} & \underline{0.2141} & \underline{0.1596} & 0.2455 & 0.3437 \\
    & \textbf{FairDDA} & \textbf{0.2152} & \textbf{0.1602} & \textbf{0.2129} & \textbf{0.3039} \\
    \hline
    \end{tabular}}
\end{table}

To evaluate the recommendation utility and fairness of our method in scenarios involving multi-class sensitive attributes, we conducted experiments on the ML-1M dataset using the occupation attribute, which comprises 21 distinct categories. The experimental results are shown in Table \ref{tab:4}, with a detailed analysis provided below:

\begin{itemize}[leftmargin=*]
\item \textbf{FairDDA} maintains a clear advantage in this setting, achieving substantial fairness improvements without compromising recommendation quality. The superiority can be attributed to the combined effect of data augmentation and debiasing learning. The two augmentation strategies effectively cut off the propagation of sensitive information in the graph structure at both the data and representation-construction levels, thereby reducing its influence during model training. On the basis, the debiasing learning further refines the learned user representations, removing residual sensitive information while retaining features relevant to user preferences.

\item Among the baseline methods, \textbf{FairCoRe} stands out by achieving an optimal balance between accuracy and fairness. Although \textbf{FairIB} improves fairness, it suffers from a clear drop in recommendation accuracy compared to the base method, LightGCN. \textbf{FairMI}, in contrast, preserves a high level of accuracy but yields only limited gains in fairness. Other methods, such as \textbf{FairRec}, \textbf{FairGo}, and \textbf{FairGNN}, suffer from a pronounced decline in accuracy with negligible fairness improvement. This is because these methods primarily rely on debiasing strategies during the representation learning, while neglecting the impact of data-level bias on fair representation learning. As a result, sensitive attributes remain in the learned representations, substantially undermining the fairness enhancement.
\end{itemize}

\end{document}